\documentclass[12pt]{article}

\usepackage[left=2.8cm,right=2.8cm,top=2.8cm,bottom=2.8cm]{geometry}

\usepackage[colorlinks=true,linkcolor=black,citecolor=black,urlcolor=blue,filecolor=black]{hyperref}

\usepackage{amsmath}
\usepackage{amsfonts}
\usepackage[nosort]{cite}

 \csname
@addtoreset\endcsname{equation}{section}

\newcommand{\supsuboverbrace}[2][]{%
  {\everymath{\scriptstyle}%
   \underbrace{\scriptstyle#2}_{#1}}%
}

\newcommand{\supsuboverbrac}[2][]{%
  {\everymath{\scriptstyle}%
   \overbrace{\scriptstyle#2}^{#1}}%
}

\def\a{\alpha}
\def\b{\beta}
\def\g{\gamma}
\def\G{\Gamma}
\def\d{\delta}
\def\D{\Delta}
\def\e{\epsilon}
\def\ve{\varepsilon}

\def\h{\eta}
\def\th{\theta}
\def\Th{\Theta}

\def\l{\lambda}
\def\L{\Lambda}
\def\m{\mu}
\def\n{\nu}
\def\x{\xi}

\def\P{\Pi}

\def\r{\rho}

\def\s{\sigma}

\def\vf{\varphi}

\def\o{\omega}
\def\O{\Omega}

\def\cC{{\cal C}}

\def\cG{{\cal G}}
\def\cH{{\cal H}}

\def\cL{{\cal L}}

\def\cN{{\cal N}}
\def\cO{{\cal O}}

\def\cT{{\cal T}}

\def\be{\begin{equation}}
\def\ee{\end{equation}}
\def\bea{\begin{eqnarray}}
\def\eea{\end{eqnarray}}
\def\ba{\begin{array}}
\def\ea{\end{array}}

\def\nn{\nonumber}

\def\nd{\nabla\!\cdot}

\def\12{\frac{1}{2}}
\def\pr{\partial}
\def\prd{\partial \cdot}

\newcommand{\und}[1]{\underline{#1}}

\begin{document}

%\begin{flushright}
%\begin{tabular}{c}
%Preprint number
%\end{tabular}
%\end{flushright}

\vspace{30pt}

\begin{center}

%%%%%%%%%%%%%%%%%%%%%%%%%%%%%%%%%%%%%%%%%%%%%%%%%%%%%%%%%%%%%%%%%%%%

{\Large\sc Higher-spin charges in Hamiltonian form \\[10pt]
I. Bose fields} 

%%%%%%%%%%%%%%%%%%%%%%%%%%%%%%%%%%%%%%%%%%%%%%%%%%%%%%%%%%%%%%%%%%%%

\vspace{25pt}
{\sc A.~Campoleoni${}^{\, a,}$\footnote{Postdoctoral Researcher of the Fund for Scientific Research-FNRS Belgium.}, M.~Henneaux${}^{\, a}$, S.~H\"ortner${}^{\, b}$ and A.~Leonard${}^{\, a,}$\footnote{Research Fellow of the Fund for Scientific Research-FNRS Belgium.}}

\vspace{10pt}
{${}^a$\sl\small
Universit{\'e} Libre de Bruxelles\\
and International Solvay Institutes\\
ULB-Campus Plaine CP231\\
1050 Brussels,\ Belgium
\vspace{10pt}

${}^b$\sl \small Centro de Estudios Cient{\'\i}ficos (CECs)\\ Casilla 1469, Valdivia, Chile\\
\vspace{10pt}

{\it andrea.campoleoni@ulb.ac.be, henneaux@ulb.ac.be,\\ hortner@cecs.cl, amaury.leonard@ulb.ac.be}

}

%%%%%%%%%%%%%%%%%%%%%%%%%%%%%%%%%%%%%%%%%%%%%%%
\vspace{40pt} {\sc\large Abstract} \end{center}

\noindent
We study asymptotic charges for symmetric massless higher-spin fields on Anti de Sitter backgrounds of arbitrary dimension within the canonical formalism. We first analyse in detail the spin-3 example: we cast Fronsdal's action in Hamiltonian form, we derive the charges and we propose boundary conditions on the canonical variables that secure their finiteness. We then extend the computation of charges and the characterisation of boundary conditions to arbitrary spin.

%%%%%%%%%%%%%%%%%%%%%%%%%%%%%%%%%%%%%%%%%%%%%%%

\newpage

%%%%%%%%%%%%%%%%%%%%%%%%%%%%%%%%%%%%%%%%%%%%%%%%%%%%%%%%%%%%%%%%%%%%

\tableofcontents

%\newpage

%%%%%%%%%%%%%%%%%%%%%%%%%%%%%%%%%%%%%%%%%%%%%%%%%%%%%%%%%%%%%%%%%%%%%
\section{Introduction}\label{sec:intro}
%%%%%%%%%%%%%%%%%%%%%%%%%%%%%%%%%%%%%%%%%%%%%%%%%%%%%%%%%%%%%%%%%%%%%

In theories with a gauge freedom, conserved quantities associated with the gauge symmetry are given by surface integrals.  For instance, the total energy and angular momentum in general relativity, or color charges in Yang-Mills theories,  are given by surface charges (see e.g.~\cite{ADM,Regge,AD1,AD2}).  These charges  play an important role in the study of the dynamics  and can be derived employing a variety of methods  (see e.g.~\cite{review-charges} for a review with references to the original literature).

A notable class of gauge theories --~whose study is receiving a renewed interested in recent years in view of their connections with string theory and the AdS/CFT correspondence~-- is given by higher-spin theories (see e.g.\ the  reviews \cite{review-old,Bekaert:2005vh,review-Vasiliev,review-strings,review-Giombi}). These models involve massless particles with spin $s>2$, actually an infinite tower of them if the dimension of space-time is bigger than three (see e.g.~\cite{review-interactions}). Their covariant description requires gauge symmetries \cite{Fronsdal,Curtright} which generalise the diffeomorphism covariance of general relativity. Consequently, they bring new surface charges in the analysis of the dynamics. 

These ``higher-spin'' charges have been considered in constant-curvature spaces in \cite{HScharges} and \cite{unfolded-charges1,unfolded-charges2}. The first reference discusses charges in arbitrary space-time dimensions; it relies on Fronsdal's formulation of the dynamics and on the covariant methods of \cite{covariant-charges1,covariant-charges2}. The following references investigate charges in four space-time dimensions within Vasiliev's unfolded formulation (see e.g.~\cite{review-old,Bekaert:2005vh,review-Vasiliev}). In three space-time dimensions higher-spin charges have also been derived from the Chern-Simons formulation of higher-spin models in both constant-curvature \cite{HR,CS3,deSitter} and flat backgrounds \cite{flat1,flat2}.

The aim of the present paper is to study higher-spin charges for symmetric massless bosonic fields within the canonical formalism, along the lines developed for general relativity in \cite{Regge,AD1,HT,BrownHenneaux,AdS-generic}. A companion paper is devoted to symmetric massless fermionic fields \cite{charges-fermions}. On the one hand, our work fits within a wider program of improving the grasp over the Hamiltonian description of bosonic higher-spin theories \cite{Hframe,Hbose,prepotentials}. On the other hand, presenting higher-spin charges in Hamiltonian form may help testing the expected thermodynamical properties of the proposed black hole solutions of  Vasiliev's equations \cite{4D-BH1,4D-BH2}, in analogy with what happened for higher-spin black holes in three space-time dimensions \cite{BH1,BH2,BH3,BH4}. In that case, the distinction between dynamical variables and Lagrange multipliers has been indeed instrumental in introducing the chemical potentials associated to the higher-spin charges carried by generalised black holes \cite{BH4}. The exact solutions of \cite{4D-BH1,4D-BH2} are also expected to be endowed with higher-spin charges \cite{unfolded-charges2},\footnote{Conserved higher-spin charges have been argued to clash with interactions by several authors. A non-vanishing cosmological constant allows however to bypass these difficulties in any space-time dimension by acting as an infrared regulator (see e.g.~\cite{review-interactions} for a review).} but the study of their thermodynamics is still in its infancy (although entropy and thermal equilibrium should be the most striking signatures of a ``generalised'' black hole). Quite generally, the canonical formalism provides a solid framework for analysing conserved charges and asymptotic symmetries. One virtue of the Hamiltonian derivation is indeed that the charges are clearly related to the corresponding symmetry. The charges play the dual role of being conserved through Noether theorem, but also of generating the associated symmetry through the Poisson bracket. This follows from the action principle. For these reasons, our work may also be useful to further develop the understanding of holographic scenarios involving higher-spin fields \cite{review-Giombi}.

In four space-time dimensions or higher, several holographic conjectures indeed anticipated a careful analysis of the Poisson algebra of AdS higher-spin charges, which defines the asymptotic symmetries of the bulk theory to be matched with the global symmetries of the boundary dual. The study of asymptotic symmetries in three dimensions \cite{HR,CS3,GH,Wlambda} proved instead crucial to trigger the development of a higher-spin AdS$_3$/CFT$_2$ correspondence \cite{review-3D}. Similarly, asymptotic symmetries played an important role in establishing new links between higher-spin theories and string theory, via the embedding of the previous holographic correspondences in stringy scenarios \cite{ABJ-HS,stringy3D} (see also sect.~6.5 of \cite{review-strings} for a review). The latter studies strongly rely on Vasiliev's equations in $d = 3, 4$ space-time dimensions, so that the tools we develop in this paper could also help in studying possible generalisations thereof.

More precisely, we compute surface charges starting from the rewriting in Hamiltonian form of Fronsdal's action on Anti de Sitter (AdS) backgrounds of arbitrary dimension \cite{fronsdal-AdS,fronsdal-AdS-D,Hbose}. This is the action describing the \emph{free} dynamics of higher-spin particles; nevertheless we expect that the rather compact final expression for the charges --~displayed in \eqref{Q-final} for the spin-3 example and in \eqref{qfin} for the generic case~-- will continue to apply even in the full non-linear theory, at least in some regimes and for a relevant class of solutions. This expectation, further discussed in sect.~\ref{sec:conclusions}, is supported by several examples of charges linear in the fields appearing in gravitational theories. It also agrees  with a previous analysis of asymptotic symmetries of three-dimensional higher-spin gauge theories \cite{metric3D} based on the perturbative reconstruction of the interacting theory within  Fronsdal's formulation \cite{metric-like,metric-like2}. Our setup is therefore close to the Lagrangian derivation of higher-spin charges of \cite{HScharges}. We compare explicitly the two frameworks in the spin-3 case at the end of sect.~\ref{sec:spin3-charges}, while our analysis also proceeds further by showing how imposing boundary conditions greatly simplifies the form of the charges at spatial infinity. This additional step allows a direct comparison between the surface charges of the bulk theory and the global charges of the putative boundary dual theories, which fits within the proposed AdS/CFT correspondences.  

The paper is organised as follows: in order to highlight the key points of our analysis of higher-spin charges, we begin by discussing in sect.~\ref{sec:spin3} the simplest example given by a spin-3 field. We detail the Hamiltonian description of the free dynamics and we provide boundary conditions on the canonical variables that secure finiteness of charges. In sect.~\ref{sec:spins} we move to arbitrary spin: we first identify the Hamiltonian constraints that generate Fronsdal's gauge transformations and then the associated charges. We eventually present boundary conditions inspired by the behaviour at spatial infinity of the solutions of the free equations of motion (recalled in Appendix~\ref{app:boundary}) and we verify that they give finite charges. We conclude summarising our results and discussing their expected regime of applicability. Other appendices provide a summary of our conventions (Appendix~\ref{app:conventions}) and more details on various results used in the main text (appendices \ref{app:fronsdal}, \ref{app:identities} and \ref{app:charges}).

%%%%%%%%%%%%%%%%%%%%%%%%%%%%%%%%%%%%%%%%%%%%%%%%%%%%%%%%%%%%%%%%%%%%%
\section{Spin-3 example}\label{sec:spin3}
%%%%%%%%%%%%%%%%%%%%%%%%%%%%%%%%%%%%%%%%%%%%%%%%%%%%%%%%%%%%%%%%%%%%%

To compute surface charges within the canonical formalism, we first rewrite in Hamiltonian form the Fronsdal action for a spin-3 field on an Anti de Sitter background of dimension $d$. The charges are then identified with the boundary terms that enter the generator of gauge transformations and we propose boundary conditions on fields and deformation parameters that secure their finiteness. The final expression for the charges is given in sect.~\ref{sec:spin3-charges}, where we also compare our outcome with other results in literature.

%%%%%%%%%%%%%%%%%%%%%%%%%%%%%%%%%%%%%%%%%%%%%%%%%%%%%%%%%%%%%%%%%%%%%
\subsection{Hamiltonian and constraints}\label{sec:spin3-H}
%%%%%%%%%%%%%%%%%%%%%%%%%%%%%%%%%%%%%%%%%%%%%%%%%%%%%%%%%%%%%%%%%%%%%

We begin with the manifestly covariant Fronsdal action \cite{fronsdal-AdS,fronsdal-AdS-D}
\be \label{fronsdal-action}
\begin{split}
S = \int\! d^{\,d}x\, \sqrt{-\bar{g}}\, \bigg\{ & -\frac{1}{2}\, \bar{\nabla}_{\!\m} \vf_{\n\r\s} \bar{\nabla}^{\m} \vf^{\n\r\s} + \frac{3}{2}\, \bar{\nabla}_{\!\m} \vf_{\n\r\s} \bar{\nabla}^{\n} \vf^{\m\r\s} - 3\, \bar{\nabla}\!\cdot \vf^{\m\n} \bar{\nabla}_{\m} \vf_{\n} \\
& + \frac{3}{2}\, \bar{\nabla}_{\!\m} \vf_{\n} \bar{\nabla}^{\m} \vf^{\n} + \frac{3}{4} \left( \bar{\nabla}\!\cdot \vf \right)^2 - \frac{2d}{L^2} \left( \vf_{\m\n\r} \vf^{\m\n\r} - \frac{3}{2}\, \vf_\m \vf^\m \right)\! \bigg\} \, .
\end{split}
\ee
Here $\bar{\nabla}$ denotes the AdS covariant derivative, $L$ is the AdS radius\footnote{All results of this subsection apply also to de Sitter provided that one maps $L \to i L$.} and omitted indices signal a trace, e.g.\ $\vf_\m = \vf_{\m\l}{}^\l$. If one parameterises the AdS$_d$ background with static coordinates
\be \label{AdS}
ds^2 = - f^2(x^k) dt^2 + g_{ij}(x^k) dx^i dx^j \, ,
\ee 
one finds that the terms in the Lagrangian with two time derivatives are
\be \label{2dots}
\cL = \sqrt{g} \left\{ \frac{f^{-1}}{2}\, \dot{\vf}^{ijk} \left( \dot{\vf}_{ijk} - 3\, g_{ij} \dot{\vf}_k \right) + \frac{f^{-7}}{4} \left( \dot{\vf}_{000} - 3f^{2} \dot{\vf}_{0} \right)^2 + \cdots \right\} ,
\ee
where Latin indices take values along spatial directions, omitted indices denote from now on a spatial trace and $g$ is the determinant of the spatial metric. Integrating by parts one can also eliminate all time derivatives acting on $\vf_{00i}$ and on the remaining contributions in $\vf_{0ij}$.
In analogy with linearised gravity, the spatial components of the covariant field thus play the role of canonical variables. The novelty is that this role is played also by the combination $(\vf_{000} - 3f^{2} \vf_{0})$, while the remaining components of the symmetric tensor are Lagrange multipliers which enforce first class constraints. That $(\vf_{000} - 3f^{2} \vf_{0})$ is a dynamical variable, in much the same way as the purely spatial components $\vf_{ijk}$, can be also inferred from its gauge transformation, which does not involve time derivatives (see the discussion around \eqref{lambda}).

It is convenient to perform the  redefinitions
\be \label{defN}
\a \equiv f^{-3} \vf_{000} - 3f^{-1} \vf_{0i}{}^i \, , \qquad
\cN_i \equiv f^{-1} \vf_{00i} \, , \qquad
N_{ij} \equiv \vf_{0ij} \, .
\ee
Introducing then the conjugate momenta to $\vf_{ijk}$ and $\alpha$, 
\be \label{Pijk}
\begin{split}
\P^{ijk} \equiv \frac{\d \cL}{\d \vf_{ijk}} & = \frac{\sqrt{g}}{f}\, \bigg\{ \dot{\vf}^{ijk} -3\, g^{(ij} \dot{\vf}^{k)} - 3 \, \nabla^{(i} N^{jk)} + 3\,g^{(ij|}\! \left(\, 2\, \nd N^{|k)} + \pr^{|k)} N \,\right) \\
& + \frac{3f}{2}\, g^{(ij} g^{k)l}\! \left(\, \pr_l \a - \G^0{}_{0l} \a \,\right)  \! \bigg\} \, ,
\end{split}
\ee
\be \label{Pt}
\tilde{\P} \equiv \frac{\d \cL}{\d \a} = \frac{\sqrt{g}}{2f} \left\{ \dot{\a} + 3\, \nd \cN \right\} ,
\ee
one can equivalently rewrite the action \eqref{fronsdal-action} in Hamiltonian form as\footnote{The rewriting of Fronsdal's action in Hamiltonian form has been previously discussed in \cite{Hbose}, relying however on the Poincar\'e parameterisation of the AdS metric. See also \cite{Hframe} for another Hamiltonian description of the free higher-spin dynamics on AdS.}
\be \label{action}
S[ \vf_{ijk}, \alpha, \P^{ijk}, \tilde{\P}, \cN^i , N^{ij}]= \int\! d^{\,d}x\, \Big\{ \P^{ijk} \dot{\vf}_{ijk} + \tilde{\P} \dot{\a} - \cH - \cN^i \cC_i - N^{ij} \cC_{ij} \Big\} \, ,
\ee
where $\cH$, $\cC_i$ and $\cC_{ij}$ are functions only of $\vf_{ijk}$, $\a$ and of their conjugate momenta $\P^{ijk}$, $\tilde{\P}$. 
Here $\nabla$ denotes the Levi-Civita connection for the spatial metric $g_{ij}$, while the ``extrinsic'' Christoffel symbol depends on $g_{00}$ as
\be \label{christoffel}
\G^0{}_{0i} = f^{-1} \pr_i f \, .
\ee
Parentheses denote a symmetrisation of the indices they enclose, and dividing by the number of terms in the sum is understood.
While \eqref{Pijk} and \eqref{Pt} hold for any static metric, the rewriting of the Fronsdal action \eqref{fronsdal-action} in the form \eqref{action} requires that the metric \eqref{AdS} be of constant curvature (see Appendix~\ref{app:fronsdal-3} for details).

The Hamiltonian in \eqref{action} reads explicitly
\begin{align}
& \cH = f \bigg\{ \frac{1}{\sqrt{g}} \! \left[\, \frac{1}{2}\, \P^{ijk}\! \left( \P_{ijk} - \frac{3}{d}\, g_{ij} \P_k \right) + \tilde{\P}^2 \,\right] + \frac{3}{2d}\, \P^i\! \left( \pr_i \a - \G^0{}_{0i}\a \right) \nn \\[5pt]
& + \sqrt{g}\, \bigg[\, \frac{1}{2}\, \nabla_{\!i} \vf_{jkl} \nabla^{i} \vf^{jkl} - \frac{3}{2}\, \nabla_{\!i} \vf_{jkl} \nabla^{j} \vf^{ikl} + 3\, \nabla\!\cdot \vf^{ij} \nabla_{\!i} \vf_{j} - \frac{3}{2}\, \nabla_{\!i} \vf_j \nabla^i \vf^j - \frac{3}{4} \left(\nabla\!\cdot\vf\right)^2 \nn \\[5pt]
&\qquad\ + \frac{2d}{L^2}\! \left( \vf_{ijk} \vf^{ijk} - \frac{3}{2}\, 
\vf_i \vf^i \right) + \G^0{}_{0i}\! \left( 3\,\vf^{ijk} \nabla_{\!j}\vf_k - \frac{3}{2}\, \vf^i \nabla\cdot \vf - \frac{9}{4}\,  \G^0{}_{0j} \vf^i \vf^j \right) \nn \\[5pt]
&\qquad\ + \frac{5d-3}{8d}\, \pr_i \a\, \pr^i \a + \frac{d}{L^2}\, \a^2 + \frac{3(d+1)}{8d}\, g^{ij} \G^0{}_{0i}\! \left( 2\,\a \pr_j \a - \G^0{}_{0j} \a^2 \right) \bigg]\bigg\} \, . \label{H}
\end{align}
The constraints are a generalisation of the Hamiltonian constraint in linearised gravity (but note here the dependence on the additional momentum $\tilde{\P}$),
\be \label{Ci}
\begin{split}
\cC_i & = 3\, \bigg\{ \pr_{i} \tilde{\P} - \frac{\sqrt{g}}{2} \bigg[\, 2 \left( \D \vf_i - \nd\nd \vf_i \right) + \nabla_i \nd \vf \\
& - \frac{4d-1}{L^2}\, \vf_i - 3\, \G^0{}_{0j}\nabla_{\!i} \vf^j + 3\, \G^0{}_{0i}\G^0{}_{0j} \vf^j \bigg] \bigg\} \, ,
\end{split}
\ee
and a generalisation of the constraint that generates spatial diffeomorphisms,
\be \label{Cij}
\begin{split}
\cC_{ij} & = -\, 3\, \bigg\{ \nd \P_{ij} + \frac{\sqrt{g}}{2}\, g_{ij}\! \left( \D - \frac{d-1}{L^2} \right) \a \\
& + \sqrt{g} \left[\, \G^0{}_{0(i} \nabla_{\!j)} \a - \G^0{}_{0i} \G^0{}_{0j} \a + \frac{1}{2}\, g_{ij} g^{kl} \G^0{}_{0k} \left( \pr_l \a - \G^0{}_{0l} \a \right) \right] \!\bigg\} \, .
\end{split}
\ee
One can verify the absence of secondary constraints and that \eqref{Ci} and \eqref{Cij} are of first class. A simple way to convince oneself of these statements is to check that they provide the correct counting of local degrees of freedom (see e.g.\ \S~1.4.2 of \cite{Hbook}):
\be
\#\, \textrm{d.o.f.} = \underbrace{\frac{(d+1)!}{3!(d-2)!} + 1}_{\textrm{canonical variables/2}} - \underbrace{\left[ \frac{(d-1)d}{2} + (d-1) \right]}_{\textrm{first-class constraints}} = \frac{(d-3)(d-2)(d+2)}{3!} \, .
\ee
The right-hand side is the dimension of a representation of $so(d-2)$ labelled by a single-row Young tableau with three boxes, as it is appropriate to describe a massless spin-3 particle in $d$ space-time dimensions.

%%%%%%%%%%%%%%%%%%%%%%%%%%%%%%%%%%%%%%%%%%%%%%%%%%%%%%%%%%%%%%%%%%%%%
\subsection{Gauge transformations}\label{sec:spin3-gauge}
%%%%%%%%%%%%%%%%%%%%%%%%%%%%%%%%%%%%%%%%%%%%%%%%%%%%%%%%%%%%%%%%%%%%%

Being of first class, the constraints $\cC_i = 0$ and $\cC_{ij} = 0$ generate gauge transformations. These correspond to the variation
\be \label{cov-gauge}
\d \vf_{\m\n\r} = 3\,\bar{\nabla}_{\!(\m} \L_{\n\r)} 
\ee
that, in the covariant language, leaves the action \eqref{fronsdal-action} invariant provided that $\L^{\m\n}$ be traceless.

In the canonical formalism, the generator of gauge transformations is 
\be \label{G}
\cG[\xi^{ij},\l^i] = \int d^{\,d-1}x \left( \x^{ij} \cC_{ij} + 
\l^i \cC_i \right) + Q_1[\x^{ij}] + Q_2[\l^i] \, ,
\ee
where $Q_1$ and $Q_2$ are the boundary terms that one has to add in order that $\cG[\x,\l]$ admit well defined functional derivatives, i.e.\ that its variation be again a bulk integral:
\be \label{deltaG}
\d \cG = \int d^{\,d-1}x \left( A_{ijk} \d \P^{ijk} + B^{ijk} \d \vf_{ijk} + C \d \tilde{\P} + D \d\a \right) .
\ee
From \eqref{deltaG} one can read the gauge transformations of the canonical variables as
\begin{subequations} \label{gauge-canonical}
\begin{alignat}{3}
\d \vf_{ijk} & = \{ \vf_{ijk} , \cG[\xi,\l] \} = A_{ijk} \, , \qquad & \d \P^{ijk} & = \{ \P^{ijk} , \cG[\xi,\l] \} = - B^{ijk} \, , \\[5pt]
\d \a & = \{ \a , \cG[\x,\l] \} = C \, , \qquad & \d \tilde{\P} & = \{ \tilde{\P} , \cG[\x,\l] \} = - D \, .
\end{alignat}
\end{subequations}
The boundary terms $Q_1[\x]$ and $Q_2[\l]$ give the asymptotic charges (see e.g.~\cite{Regge,benguria-cordero}). Sensible boundary conditions on the canonical variables must then be chosen such that the charges be finite when evaluated on deformation parameters which generate transformations preserving the given boundary conditions. 

Inserting the definitions \eqref{Ci} and \eqref{Cij} of the constraints in \eqref{G} and taking into account that the background has constant curvature, one obtains
\begin{subequations}
\begin{align}
A_{ijk} & = 3\, \nabla_{\!(i} \x_{jk)} \, , \label{Aijk} \\[5pt]
B^{ijk} & = 3 \sqrt{g} \left\{ \nabla^{(i} \nabla^j \l^{k)}\! - g^{(ij|}\! \left[ \left( \D - \frac{2(d-1)}{L^2} \right)\! \l^{|k)} + \frac{1}{2} \left(\nabla^{|k)}\! + 3\, \G^0{}_{0}{}^{|k)} \right)\! \nd \l \right]\! \right\} , \label{Bijk} \\[5pt]
C & = -\, 3\, \nd \l \, , \label{C} \\[5pt]
D & = -\, \frac{3}{2}\sqrt{g} \left[ \left( \D - \frac{2d}{L^2} \right)\! \x - \G^0{}_{0i}\! \left( 2 \nd \x^i + \pr^i \x \right) \right] . \label{D}
\end{align}
\end{subequations}
The boundary terms generated by the integrations by parts putting the variation of $\cG[\x,\l]$ in the form \eqref{deltaG} must be cancelled by the variations $\d Q_1$ and $\d Q_2$ of the charges. Being linear in the fields, these variations are integrable and yield:\footnote{Here $d^{d-2}S_i \equiv d^{d-2} x\, \hat{n}_i$, where $\hat{n}_i$ and $d^{d-2} x$ are respectively the normal and the product of differentials of the coordinates on the $d-2$ sphere at infinity (e.g.\ $d^{2} x = d\th d\phi$ for $d=4$, so that $d^2S_i$ does not include the determinant of the intrinsic metric that appears in the full volume element).}
\begin{subequations} \label{spin3-charges}
\begin{align}
Q_1[\x^{ij}] & = 3\! \int\! d^{\,d-2}S_i \left\{ \x_{jk} \P^{ijk} + \frac{\sqrt{g}}{2} \left[\, \x \nabla^i \a - \a \nabla^i \x + \G^0{}_{0j}\! \left( 2\,\x^{ij} + g^{ij} \x \right)\! \a \,\right] \right\} , \label{Q1} \\[10pt]
Q_2[\l^i] & = 3\! \int\! d^{\,d-2}S_i\, \bigg\{ - \l^i \tilde{\P} + \frac{\sqrt{g}}{2} \left[\, 2\, \l_j \nabla^i \vf^j - 2\, \l_j \nd \vf^{ij} + \l^i \nd \vf \right. \nn \\
& \phantom{= 3\! \int\! d^{\,d-2}S_i\, \bigg\{}\, \left. - \, 2\, \vf_j \nabla^i \l^j + 2\, \vf^{ijk} \nabla_{\!j} \l_k - \nd \l \vf^i - 3\, \G^0{}_{0j} \l^i \vf^j \,\right]\! \bigg\} \, . \label{Q2}
\end{align}
\end{subequations}
In presenting $Q_1$ and $Q_2$ we also adjusted the integration constants so that the charges vanish for the zero solution.

Expanding \eqref{cov-gauge} in components and comparing with \eqref{Aijk} and \eqref{C}, one can also identify the deformation parameter $\x^{ij}$ with the spatial components of the covariant gauge parameter, while $\l^i$ is related to $\L^{\m\n}$ by
\be \label{lambda}
\l^i = -\, 2 f\, \L^{0i} \, .
\ee
The remaining component of the covariant gauge parameter, $\L^{00}$, is proportional to $g_{ij} \x^{ij}$ thanks to the Fronsdal constraint $g_{\m\n} \L^{\m\n} = 0$ that allows the cancellation of time derivatives in the gauge variation of $(\vf_{000} - 3f^{2} \vf_{0})$. The other components of \eqref{cov-gauge} give the gauge transformations of the Lagrange multipliers:
\begin{subequations}
\begin{align}
\d N_{ij} & = \dot{\x}_{ij} + f \left( \nabla_{\!(i} \l_{j)} - \G^0{}_{0(i}\l_{j)} \right) , \\[5pt]
\d \cN_i & = \dot{\l}_i + f \left( \pr_i \x - 2\,\G^0{}_{0i} \x - 2\, \G^0{}_{0j} \x_i{}^j \right) .
\end{align}
\end{subequations}
Substituting the previous variations in the definitions of the momenta one can finally check the consistency of \eqref{Bijk} and \eqref{D} with Fronsdal's gauge transformations.
Notice that time derivatives only appear in the gauge transformations of Lagrange multipliers, in agreement with the general results discussed e.g.\ in \S~3.2.2 of \cite{Hbook}.  This confirms once more our splitting of the covariant field into canonical variables and Lagrange multipliers. 

%%%%%%%%%%%%%%%%%%%%%%%%%%%%%%%%%%%%%%%%%%%%%%%%%%%%%%%%%%%%%%%%%%%%%
\subsection{Boundary conditions}\label{sec:spin3-bnd}
%%%%%%%%%%%%%%%%%%%%%%%%%%%%%%%%%%%%%%%%%%%%%%%%%%%%%%%%%%%%%%%%%%%%%

We now have to set boundary conditions on the canonical variables and restrict the deformation parameters to those that generate gauge transformations preserving them. Since we deal with the linearised theory, we can actually fully specify the space of solutions of the equations of motion. We then extract boundary conditions from the behaviour at spatial infinity of the solutions in a convenient gauge, with the expectation that the regime of applicability of both the charges \eqref{spin3-charges} and the following fall-off conditions will extend even beyond the linearised regime (see sect.~\ref{sec:conclusions} for an ampler discussion of this strategy).

In Appendix~\ref{app:boundary} we recall the falloff at the boundary of the two branches of solutions of the second-order equations of motion imposed by the covariant action principle. In a coordinate system in which the AdS metric reads
\be \label{poincare}
ds^2 = \frac{dr^2}{r^2} + r^2\, \h_{IJ} dx^Idx^J \, ,
\ee
the solutions in the \emph{subleading branch} behave at spatial infinity ($r \to \infty$) as 
\begin{subequations} \label{boundary}
\begin{align}
\vf_{IJK} & = r^{3-d}\, \cT_{IJK}(x^M) + \cO(r^{1-d}) \, , \label{dvIJK} \\[5pt]
\vf_{rIJ} & = \cO(r^{-d}) \, , \label{dvrIJ} \\[5pt]
\vf_{rrI} & = \cO(r^{-d-3}) \, , \label{dvrrI} \\[5pt]
\vf_{rrr} & = \cO(r^{-d-6}) \, . \label{dvrrr}
\end{align}
\end{subequations}
We remark that capital Latin indices denote all directions which are transverse to the radial one (including time) and that here and in the following we set the AdS radius to $L=1$. The field equations further impose that $\cT_{IJK}$ be conserved and traceless:
\be \label{current}
\pr^K \cT_{IJK} = \h^{JK} \cT_{IJK} = 0 \, . 
\ee
We take \eqref{boundary} and \eqref{current} as the definition of our boundary conditions.

In the case of spin $2$ included in the discussion of sect.~\ref{sec:bndH}, for which the complete theory is known in closed form (AdS gravity), the boundary conditions generally considered in the literature agree with the behaviour of the solutions in the subleading branch \cite{HT,BrownHenneaux,AdS-generic}. Since in this case finiteness of the charges and consistency have been completely checked, we also adopt here boundary conditions defined by the subleading branch, which is the direct generalisation of these previous works.  It would be of interest to extend the analysis to more general asymptotics, as done for scalar fields in \cite{scalar1,scalar2,scalar3}, but we leave this question for future work. Furthermore, for $d=3$ the boundary conditions \eqref{boundary} agree with those in eq.~(3.9) of \cite{metric3D}, which have been derived from the Chern-Simons formulation of the full interacting theory of a spin-3 field with gravity.\footnote{Actually in \cite{metric3D} the trace of $\vf_{rIJ}$ is even $\cO(r^{-d-2})$ for the following reason: in the interacting theory --~at least in $d=3$~-- the fall-off conditions \eqref{boundary} are not preserved by asymptotic symmetries unless the fields also satisfy the  equations of motion up to an order in $r$ which depends on the highest spin appearing in the model. As we shall see, these additional specifications are anyway irrelevant to compute charges.}

From \eqref{boundary} one directly gets the behaviour at spatial infinity of half of the canonical variables. Denoting the coordinates that parameterise the $d-2$ sphere at infinity by Greek letters from the beginning of the alphabet, one obtains
\begin{subequations} \label{bnd-q}
\begin{align}
\vf_{\a\b\g} & = r^{3-d}\, \cT_{\a\b\g} + \cO(r^{1-d}) \, , \qquad \a = -\frac{2}{r^d}\, \cT_{000} + \cO(r^{-d-2})\, , \\[10pt]
\vf_{r\a\b} & = \cO(r^{-d}) \, , \qquad \vf_{rr\a} = \cO(r^{-d-3}) \, , \qquad \vf_{rrr} = \cO(r^{-d-6}) \, .
\end{align}
\end{subequations}
To fix the behaviour of $\a$ we used the trace constraint on $\cT_{IJK}$, which implies $\d^{\a\b}\cT_{0\a\b} = \cT_{000}$. The boundary conditions on the momenta follow from the substitution of \eqref{bnd-q} in \eqref{Pijk} and \eqref{Pt}:
\begin{subequations} \label{bnd-p}
\begin{align}
\P^{r\a\b} & = \frac{d+1}{r^4}\, \cT_0{}^{\a\b} + \cO(r^{-6}) \, , \qquad\qquad \tilde{\P} = \cO(r^{-4}) \, , \\[10pt]
\P^{\a\b\g} & = \cO(r^{-7}) \, , \qquad 
\P^{rr\a} = \cO(r^{-3}) \, , \qquad
\P^{rrr} = \cO(1) \, . 
\end{align}
\end{subequations}
In the formulae above we displayed explicitly only the terms which contribute to the charges (see sect.~\ref{sec:spin3-charges}). We expressed everything in terms of the components of the conserved boundary current $\cT_{IJK}$, but $\cT_{\a\b\g}$, $\cT_{000}$ and $\cT_0{}^{\a\b}$ can be considered only as convenient labels to denote the boundary values of, respectively, $\vf_{\a\b\g}$, $\a$ and $\P^{r\a\b}$.  The covariant boundary conditions \eqref{boundary} would also fix the falloff of the Lagrange multipliers. The resulting conditions, however, would correspond to a particular choice of gauge, while --~as noticed in three space-time dimensions \cite{chemical,BH4}~-- the freedom in the choice of the Lagrange multipliers is instrumental in fitting some physically relevant solutions within the boundary conditions (see also \cite{BHregularity}). We postpone to future work a detailed analysis of this issue, especially because this freedom does not affect charges.

%%%%%%%%%%%%%%%%%%%%%%%%%%%%%%%%%%%%%%%%%%%%%%%%%%%%%%%%%%%%%%%%%%%%%
\subsection{Asymptotic symmetries}\label{sec:spin3-symm}
%%%%%%%%%%%%%%%%%%%%%%%%%%%%%%%%%%%%%%%%%%%%%%%%%%%%%%%%%%%%%%%%%%%%%

As a next step we identify asymptotic symmetries, that are the gauge transformations preserving boundary conditions. Our goal is to specify appropriate deformation parameters in the generator of gauge transformations \eqref{G}. We begin however by selecting covariant gauge transformations compatible with the fall-off conditions \eqref{boundary}, to later identify the corresponding deformation parameters using e.g.\ \eqref{lambda}.

Asymptotic symmetries contain at least the ``higher-rank isometries'' of the AdS space, i.e.\ the gauge transformations leaving the vacuum solution \mbox{$g_{\m\n} = g^{\textrm{AdS}}_{\m\n}$} and \mbox{$\vf_{\m\n\r} = 0$} invariant. These are generated by traceless AdS Killing tensors, satisfying
\be \label{exact-killing}
\bar{\nabla}_{\!(\m} \Lambda_{\n\r)} = 0 \, , \qquad\qquad \bar{g}_{\m\n} \Lambda^{\m\n} = 0 \, ,
\ee
where the latter condition is the usual Fronsdal constraint. These equations have been studied e.g.\ in \cite{killing,kill_tensors_1,HScharges}: they admit $\frac{(d-2)(d+1)(d+2)(d+3)}{12}$ independent solutions, obtained as traceless combinations of symmetrised products of AdS Killing vectors \cite{killing}. This guarantees that non-trivial asymptotic symmetries do exist. We shall also notice that --~as far as the free theory in $d > 3$ is concerned~-- asymptotic and exact Killing tensors only differ in terms that do not contribute to surface charges. It is nevertheless instructive to specify  asymptotic Killing tensors in a unified framework that applies for any $d \geq 3$.

In Appendix~\ref{app:gauge} we recall the behaviour at the spatial boundary of traceless AdS Killing tensors in the coordinates \eqref{poincare}. We assume that asymptotic Killing tensors have the same leading behaviour at $r \to \infty$, that is
\begin{subequations} \label{boundary-gauge}
\begin{align}
\Lambda^{IJ} & = \chi^{IJ}(x^K) + \sum_{n\,=\,1}^\infty r^{-2n}\, \chi_{(n)}^{IJ} (x^K)\, , \label{chi-exp} \\
\Lambda^{rI} & = r\, w^I(x^K) + \sum_{n\,=\,1}^\infty r^{1-2n}\, w_{(n)}^{I} (x^K)\, , \label{w-exp} \\
\Lambda^{rr} & = r^2\, u(x^K) + \sum_{n\,=\,1}^\infty r^{2-2n}\, u_{(n)} (x^K) \, , \label{u-exp}
\end{align}
\end{subequations}
and we now analyse the constraints imposed by the preservation of boundary conditions. Notice that $\Lambda^{rr}$ and the trace of $\L^{IJ}$ are not independent: Fronsdal's constraint gives 
\be \label{trace-poincare}
\Lambda^{rr} = - r^4\, \h_{IJ} \L^{IJ} \, .
\ee
As a result $\chi^{IJ}$ in \eqref{chi-exp} is traceless, while the trace of $\chi^{IJ}_{(n+1)}$ is proportional to $u_{(n)}$.

For any value of the space-time dimension $d$, the variation $\d \vf_{IJK}$ induced by \eqref{boundary-gauge} decays slower at spatial infinity than the boundary conditions \eqref{boundary}:
\be \label{leading-var}
\d \vf_{IJK} = 3\, r^4 \left\{ \pr_{(I} \chi_{JK)} + 2\, \h_{(IJ} w_{K)} \right\} + \cO(r^2) \, .
\ee
The expression between parentheses must therefore vanish; this requires that $\chi^{IJ}$ be a \emph{traceless conformal Killing tensor} for the Minkowski metric $\h_{IJ}$ (see e.g.~\cite{algebra}),\footnote{For simplicity we fixed the coordinates \eqref{poincare} such that the boundary metric is the flat Minkowski metric, but \eqref{kill-boundary} is invariant under conformal rescalings of the boundary metric. See e.g.~\cite{metric3D} for a discussion of this issue in the $d=3$ example, that can be extended verbatim to $d > 3$.}
\be \label{kill-boundary}
\pr_{(I} \chi_{JK)} - \frac{2}{d+1}\, \h_{(IJ\,} \prd \chi_{K)} = 0 \, , \qquad\qquad \h_{IJ} \chi^{IJ} = 0 \, ,
\ee
and that $w^I$ be fixed as
\be
w_I = -\, \frac{1}{d+1}\, \prd \chi_I \, .
\ee
As recalled in Appendix~\ref{app:identities}, when $d > 3$ the equations \eqref{kill-boundary} --~here defined in \mbox{$d-1$} dimensions~-- admit the same number of independent solutions as the equations \eqref{exact-killing}. It is therefore not surprising that their solutions can be extended to solutions of the Killing equation \eqref{exact-killing}, provided one fixes the subleading components of the gauge parameter in terms of $\chi^{IJ}$. As a result, in the linearised theory asymptotic Killing tensors coincide with traceless Killing tensors of the AdS space up to a certain order in $r$, except in $d = 3$ where the equations \eqref{kill-boundary} admit locally infinitely many solutions.

To prove the previous statement, let us look at the variations of the other components with at least one transverse index:
\begin{subequations} \label{var-with-I}
\begin{align}
\d \vf_{rIJ} & = \sum_{n\,=\,1}^\infty r^{3-2n} \left\{ -2n\, \chi^{(n)}_{IJ} - 2\, \h_{IJ} \chi^{(n)} + 2\, \pr_{(I}^{\phantom{(}} w_{J)}^{(n-1)} \right\} ,  \\
\d \vf_{rrI} & = \sum_{n\,=\,1}^\infty r^{-2n} \left\{ -4n\, w^{(n)}_I - \pr_I \chi^{(n)} \right\} .
\end{align}
\end{subequations}
The leading order must cancel for any $d$ and this requires 
\be
\chi^{(1)}_{IJ} = - \frac{1}{d+1} \left( \pr_{(I} \prd \chi_{J)} - \frac{1}{d}\, \h_{IJ} \prd\prd \chi \right) , \qquad
w^{(1)}_I = \frac{1}{4d(d+1)}\, \pr_I \prd\prd \chi \, .
\ee
Substituting in $\d \vf_{IJk}$ one discovers that the first subleading order in \eqref{leading-var} vanishes as well provided that 
\be \label{kill1}
\pr_{(I} \pr_{J} \prd \chi_{K)} - \frac{3}{2d}\, \h_{(IJ} \pr_{K)} \prd\prd \chi = 0 \, .
\ee
This identity is shown to follow from \eqref{kill-boundary} in Appendix~\ref{app:identities}.

The subleading orders can be analysed in a similar way. From \eqref{var-with-I} one sees that the boundary conditions \eqref{dvrIJ} and \eqref{dvrrI} are preserved provided that for $1 < n < \frac{d+1}{2}$ one has
\begin{subequations}
\begin{align}
\chi^{(n)}_{IJ} & = \frac{(-1)^n (d-1)!}{4^{n-1} n! (d+n-2)! (d+1)} \left(  \pr_I \pr_J - \frac{1}{d+n-1}\, \h_{IJ} \Box \right) \Box^{n-2} \prd\prd \chi , \\
w^{(n)}_I & = \frac{(-1)^{n+1} (d-1)!}{4^{n}n!(d+n-1)!(d+1)}\, \Box^{n-1} \pr_{I} \prd\prd \chi \, .
\end{align}
\end{subequations}
Computing three divergences of the conformal Killing equation \eqref{kill-boundary} one finds however
\be \label{kill3}
(d-1)\, \Box\, \prd\prd \chi = 0 \, ,
\ee
so that all $w^{(n)}_I$ with $n > 1$ and all $\chi^{(n)}_{IJ}$ with $n > 2$ vanish when $d > 1$. This also implies $u^{(n)} = 0$ for $n \geq 1$ thanks to the trace constraint \eqref{trace-poincare} and, in its turn, that the condition \eqref{dvrrr} is preserved ($\chi^{(2)}_{IJ}$ does not vanish but it is traceless).

The variation of the transverse component thus reduces to
\be \label{var-subleading}
\d \vf_{IJK} = \frac{1}{8d(d+1)}\, \pr_I \pr_J \pr_K \prd\prd \chi + \cO(r^{3-d}) \, .
\ee
The term written explicitly must vanish in all space-time dimensions except $d=3$ and, indeed, in Appendix~\ref{app:identities} we prove that \eqref{kill-boundary} implies
\be \label{kill2}
(d-3)\, \pr_I \pr_J \pr_K \prd\prd \chi = 0 \, .
\ee
In $d=3$ this identity is not available and one has two options: if one wants to solve the Killing equation \eqref{exact-killing} one has to impose the cancellation of the triple gradient of $\prd\prd\chi$ and the additional condition is satisfied only on a finite dimensional subspace of the solutions of the conformal Killing equation \eqref{kill-boundary}.  If one is instead interested only in preserving the boundary conditions \eqref{boundary}, which is the only option when the background is not exact AdS space, a shift of $\vf_{IJK}$ at $\cO(1)$ is allowed. The corresponding variation of the surface charges is at the origin of the central charge that appears in the algebra of asymptotic symmetries (see sect.~3.4 of \cite{metric3D}). 

To summarise: parameterising the AdS$_d$ background as in \eqref{poincare}, linearised\footnote{A similar analysis has been performed in $d=3$ including interactions \cite{metric3D}. These introduce a dependence on the boundary values of the fields in \eqref{final-gauge}, while asymptotic Killing tensors are still in one-to-one correspondence with the solutions of the conformal Killing equation.} covariant gauge transformations preserving the boundary conditions \eqref{boundary} are generated by
\begin{subequations} \label{final-gauge}
\begin{align}
\Lambda^{IJ} & = \chi^{IJ} - \frac{r^{-2}}{d+1} \left( \pr^{(I} \prd \chi^{J)} - \frac{1}{d}\,\h^{IJ} \prd\prd\chi \right) + \frac{r^{-4}}{8d(d+1)}\, \pr^I \pr^J \prd\prd \chi \\
& + \cO(r^{-d-3}) \, , \nn \\
\Lambda^{rI} & = - \frac{r}{d+1}\, \prd \chi^I + \frac{r^{-1}}{4d(d+1)}\, \pr^I \prd\prd \chi + \cO(r^{-d-2}) \, , \\[2pt]
\Lambda^{rr} & = \frac{r^2}{d(d+1)}\, \prd\prd\chi + \cO(r^{-d-1}) \, , 
\end{align}
\end{subequations}
where $\chi^{IJ}$ satisfies the conformal Killing equation \eqref{kill-boundary}, whose general solution is recalled in \eqref{sol-conf}.
In sect.~\ref{sec:spin3-gauge} we have seen that the deformation parameter $\xi^{ij}$ has to be identified with the spatial components of the covariant gauge parameter $\L^{\m\n}$, while the deformation parameter $\l^i$ is related to $\L^{\m\n}$ by \eqref{lambda}. Combining this information with \eqref{final-gauge}, one sees that asymptotic symmetries are generated by deformation parameters behaving as 
\begin{subequations} \label{bnd-deform}
\begin{align}
\x^{\a\b} & = \chi^{\a\b} + \cO(r^{-2}) \, , \qquad
\l^\a  = -\,2r\, \chi^{0\a} + \cO(r^{-1}) \, , \\[10pt] 
\x^{r\a} & = \cO(r) \, , \qquad 
\x^{rr} = \cO(r^2) \, , \qquad
\l^r = \cO(r^{2}) \, .
\end{align}
\end{subequations}
As in \eqref{bnd-q}, Greek letters from the beginning of the alphabet denote coordinates on the $d-2$ sphere at infinity and we specified the dependence on $\chi^{IJ}$ only in the terms that contribute to surface charges. In particular, as for gravity, when $d > 3$ the charges are insensitive to the differences between asymptotic and exact Killing tensors, corresponding to the unspecified subleading orders in \eqref{final-gauge}.

%%%%%%%%%%%%%%%%%%%%%%%%%%%%%%%%%%%%%%%%%%%%%%%%%%%%%%%%%%%%%%%%%%%%%
\subsection{Charges}\label{sec:spin3-charges}
%%%%%%%%%%%%%%%%%%%%%%%%%%%%%%%%%%%%%%%%%%%%%%%%%%%%%%%%%%%%%%%%%%%%%

Having proposed boundary conditions on both canonical variables (see \eqref{bnd-q} and \eqref{bnd-p}) and deformation parameters (see \eqref{bnd-deform}), we can finally evaluate the asymptotic charges \eqref{spin3-charges}. In the coordinates \eqref{poincare}, the normal to the $d-2$ sphere at infinity is such that $\hat{n}_r = 1$ and $\hat{n}_\a = 0$. At the boundary the charges thus simplify as
\begin{subequations} \label{Q-final}
\begin{align}
\lim_{r\to\infty}Q_1[\x^{ij}] & = 3\! \int\! d^{\,d-2}x \left\{ \x_{\a\b} \P^{r\a\b}\! + \frac{\sqrt{g}}{2}\, g^{rr} g_{\a\b}\! \left( \x^{\a\b} \pr_r \a - \a \nabla_{\!r} \x^{\a\b} + \G^0{}_{0r} \x^{\a\b} \a \right) \! \right\} , \label{Q1-final} \\[5pt]
\lim_{r\to\infty}Q_2[\l^{i}] & = 3\! \int\! d^{\,d-2}x\, \sqrt{g}\, g^{rr} g^{\a\b}\! \left\{ \l^\g \nabla_{\!r} \vf_{\a\b\g} + \l^\g \G^\d{}_{r\a} \vf_{\b\g\d} - \vf_{\a\b\g} \nabla_{\!r} \l^\g \right\} . \label{Q2-final}
\end{align}
\end{subequations}
The terms which survive in the limit give a finite contribution to the charges; one can make this manifest by substituting their boundary values so as to obtain
\begin{subequations} \label{Q-chi}
\begin{align}
\lim_{r\to\infty}Q_1 & = 3(d+1)\! \int\! d^{\,d-2}x \left( \chi^{\a\b} \cT_{0\a\b} + \chi^{00} \cT_{000} \right) , \\[5pt]
\lim_{r\to\infty}Q_2 & = 6(d+1)\! \int\! d^{\,d-2}x\, \chi^{0\a} \cT_{00\a} \, , 
\end{align}
\end{subequations}
where we used the trace constraints on both $\chi^{IJ}$ and $\cT_{IJK}$.
When one sums both charges, the result partly covariantises in the indices transverse to the radial direction:
\be \label{charge-cov}
Q \equiv \lim_{r\to\infty}\! \left( Q_1 + Q_2 \right) = 3 (d+1) \!\int\! d^{\,d-2}x\, \chi^{IJ} \cT_{0IJ} \, .
\ee
The boundary charge thus obtained is manifestly conserved: it is the spatial integral of the time component of a conserved current since
\be
J_I \equiv \chi^{JK} \cT_{IJK} \quad \Rightarrow \quad \prd J = \pr^{(I} \chi^{JK)} \cT_{IJK} + \chi^{JK} \prd \cT_{JK} = 0 \, ,
\ee
where the conservation holds thanks to \eqref{current} and \eqref{kill-boundary}. We recovered in this way the standard presentation of the global charges of the boundary theories entering the higher-spin realisations of the AdS/CFT correspondence (see e.g.\ sect.~2 of \cite{review-Giombi} for a review).

In three space-time dimensions, the rewriting \eqref{charge-cov} exhibits the chiral splitting of charges that one obtains in the Chern-Simons formulation with a suitable choice of the boundary value of the Lagrange multipliers\cite{HR,CS3,BH4}. Introducing the light-cone coordinates $x^\pm = t \pm \phi$, one obtains
\be \label{3Dcharge}
\begin{split}
Q_{d=3} & = 12 \!\int\! d\phi \left( \chi^{IJ} \cT_{+IJ} + \chi^{IJ} \cT_{-IJ} \right) = 12 \!\int\! d\phi \left( \chi^{++} \cT_{+++} + \chi^{--} \cT_{---} \right) \\
& = 12 \!\int\! d\phi \left( \chi(x^+) \cT(x^+) + \tilde{\chi}(x^-) \tilde{\cT}(x^-) \right) ,
\end{split}
\ee
where we took advantage of the form of the general solutions of \eqref{current} and \eqref{kill-boundary}, see e.g.~\eqref{sol-conf3}.
The separation in $Q_1$ and $Q_2$ does not correspond, however, to the splitting in left and right-moving components since
\begin{align}
\lim_{r\to\infty}Q_1 & = 6\! \int\! d\phi\, ( \chi + \tilde{\chi} ) ( \cT + \tilde{\cT} ) \, , \qquad
\lim_{r\to\infty}Q_2 = 6\! \int\! d\phi\, ( \chi - \tilde{\chi} ) ( \cT - \tilde{\cT} ) \, . 
\end{align}

The analysis of the linearised theory allowed us to recover the expression \eqref{3Dcharge} that also holds in the non-linear theory \cite{HR,CS3}, in line with our general expectations. Note, however, that in $d > 3$ the asymptotic symmetries \eqref{final-gauge} leave the charges \eqref{charge-cov} invariant, while in $d=3$ only a variation of \eqref{3Dcharge} which is independent of $\cT$ and $\tilde{\cT}$ is allowed. Since the variation of the charges is generated by the charges themselves \cite{Brown:1986ed} as
\be
\d_{\xi_2} Q[\x_1] = \{ Q[\xi_1] , Q[\xi_2] \} \, , 
\ee
one concludes that --~although the linearised theory suffices to identify the charges~-- their algebra does depend on interactions. As discussed in sect.~\ref{sec:conclusions}, this phenomenon is anyway not a peculiarity of higher-spin theories. For an explicit example of the correlation between algebras of charges and interactions we refer to \cite{metric3D} where, assuming the expression \eqref{3Dcharge} for spin-3 charges in $d=3$, their algebra has been computed by including the first interaction vertices in a weak-field expansion in the spin-3 field.

We conclude this section by explicitly comparing our results with the higher-spin charges derived from Fronsdal's action in \cite{HScharges} within the covariant approach of \cite{covariant-charges1,covariant-charges2}.\footnote{See also Appendix~D of \cite{covariant-charges2}, where the methods of \cite{covariant-charges1,covariant-charges2} are applied to Hamiltonian actions to discuss the equivalence between canonical and covariant derivations of surface charges.} In the latter context, the spin-3 charges are given by the integral on a closed $d-2$ surface of the \emph{on-shell} closed $(d-2)$-form
\be \label{form}
k[\L] = \frac{1}{2(d-2)!}\, k^{[\m\n]}[\L]\, \e_{\m\n\r_3 \cdots \r_{d}} dx^{\r_3} \cdots dx^{\r_{d}} \, ,
\ee
with
\be \label{spin3-covariant}
\begin{split}
k^{[\m\n]}[\L] = \sqrt{-\bar{g}}\, & \Big\{\, \vf_{\r\s}{}^{[\m} \bar{\nabla}^{\n]} \Lambda^{\r\s} + \vf_\r \bar{\nabla}^{[\m} \Lambda^{\n]\r} + \Lambda^{\r\s} \bar{\nabla}^{[\m} \vf^{\n]}{}_{\r\s} \\
& + \L_\r{}^{[\m|}\! \left(\, 2\,\bar{\nabla}^{|\n]} \vf^\r + \bar{\nabla}^\r \vf^{|\n]} - 2\,\bar{\nabla}\!\cdot \vf^{|\n]\r} \,\right) \Big\} \, .
\end{split}
\ee
Here we went back to a manifestly covariant notation as in \eqref{fronsdal}, while square brackets denote an antisymmetrisation.
The closure of $k[\L]$ --~from which conservation of the charges follows~-- requires that the traceless parameter $\Lambda^{\m\n}$ satisfies the Killing tensor equation \eqref{exact-killing}.
By integrating \eqref{form} on a $d-2$ sphere at fixed time and radial distance from the origin and using \eqref{lambda}, one obtains
\be \label{compare-charges}
\int\! d^{\,d-2}S_i\, k^{0i}[\L] = -\, \frac{1}{3} \left( Q_1[\x] + Q_2[\l] \right) + \int\! d^{\,d-2}S_i\, \sqrt{g} \left( K_1^i + K_2^i \right) . 
\ee
One thus recovers the canonical charges \eqref{spin3-charges} plus a contribution that vanishes after integration since
\be \label{k1}
K_1^i = \nabla_{\!j}\! \left[\, \l^{[i}\vf^{j]} - f^{-1}\! \left( \l^{[i}\cN^{j]} - 4\, \x^{k[i}N^{j]}{}_k \right) \right]
\ee
is a total derivative on the sphere, while
\be \label{k2}
\begin{split}
K_2^i = & - \frac{3}{f} \left( g^{ij} N^{kl} - g^{kl} N^{ij} \right) \bar{\nabla}_{\!(j} \Lambda_{kl)} + \frac{3}{f} \left( \vf^{ijk} - g^{ij} \vf^k + f^{-1} g^{ij} \cN^k \right) \bar{\nabla}_{\!(0} \Lambda_{jk)} \\
& - \frac{3}{2f^2} \left( g^{ij} \a + 2f^{-1} \left( N^{ij} + g^{ij} N \right) \right) \bar{\nabla}_{\!(0} \Lambda_{0j)} + \cN^i\, \bar{\nabla}\!\cdot \L^0
\end{split}
\ee
vanishes identically for parameters that satisfy \eqref{exact-killing}. The deformation parameters in \eqref{spin3-charges} satisfy the Killing tensor equation \eqref{exact-killing} at leading order, so that eq.~\eqref{compare-charges} implies that the covariant and canonical expressions for the charges coincide asymptotically. Similarly, the constraint on the $\L^{\m\n}$ in \eqref{spin3-covariant} can be imposed only asymptotically to capture the infinitely many conserved charges that we exhibited in three space-time dimensions (see e.g.~\cite{covariant3d} for a discussion of $3d$ gravity in the covariant context).

%%%%%%%%%%%%%%%%%%%%%%%%%%%%%%%%%%%%%%%%%%%%%%%%%%%%%%%%%%%%%%%%%%%%%
\section{Arbitrary spin}\label{sec:spins}
%%%%%%%%%%%%%%%%%%%%%%%%%%%%%%%%%%%%%%%%%%%%%%%%%%%%%%%%%%%%%%%%%%%%%

In this section we extend most of the previous results to arbitrary spin. To simplify expressions we resort to an index-free notation: we omit indices and we denote e.g.\ the \mbox{$n$-th} trace of the field $\vf_{\m_1 \cdots \m_s}$ by $\vf^{[n]}$. Symmetrised gradients and divergences are denoted by $\nabla$ and $\nabla\cdot$, while in terms which are quadratic in the fields a contraction of all free indices is understood. A summary of our conventions is presented in Appendix~\ref{app:conventions}.

%%%%%%%%%%%%%%%%%%%%%%%%%%%%%%%%%%%%%%%%%%%%%%%%%%%%%%%%%%%%%%%%%%%%%
\subsection{Constraints and gauge transformations}\label{sec:gauge}
%%%%%%%%%%%%%%%%%%%%%%%%%%%%%%%%%%%%%%%%%%%%%%%%%%%%%%%%%%%%%%%%%%%%%

In Fronsdal's formulation of the dynamics, a spin-$s$ particle is described by a symmetric tensor of rank $s$ with vanishing double trace \cite{fronsdal-AdS}. The action can be cast in the form
\begin{align}
S =\! \int & d^{\,d}x \sqrt{-\bar{g}}\, \bigg\{\! - \12\, \bar{\nabla}_{\!\m} \vf \bar{\nabla}^{\m} \vf + \frac{s}{2}\, \bar{\nabla}_{\!\m} \vf_{\n} \bar{\nabla}^\n \vf^{\m} - \binom{s}{2} \bar{\nabla}\!\cdot \vf^\m \bar{\nabla}_{\!\m} \vf^{[1]} + \12 \binom{s}{2} \bar{\nabla}_{\!\m} \vf^{[1]} \bar{\nabla}^{\m} \vf^{[1]} \nn \\
& + \frac{3}{4} \binom{s}{3} \bar{\nabla}\!\cdot \vf^{[1]} \bar{\nabla}\!\cdot \vf^{[1]} - \frac{(s-1)(d+s-3)}{L^2}\, \vf \left( \vf - \frac{1}{2}\binom{s}{2} g\, \vf^{[1]} \right)\! \bigg\} \, , \label{fronsdal-s}
\end{align}
where $\vf^{[1]}$ denotes a trace and omitted indices are understood to be contracted as e.g.\ in
\be
\bar{\nabla}_{\!\m} \vf_{\n} \bar{\nabla}^\n \vf^{\m} \equiv \bar{\nabla}_{\!\m} \vf_{\n\r_1 \cdots\, \r_{s-1}} \bar{\nabla}^\n \vf^{\m\r_1 \cdots\, \r_{s-1}} \, .
\ee 
The action \eqref{fronsdal-s} is invariant under the transformations
\be \label{fronsdal-gauge}
\d\vf = s\, \bar{\nabla} \L \, , \qquad\quad \L^{[1]} = 0 \, ,
\ee
where, as in \eqref{cov-gauge}, a symmetrisation of the indices carried by the gradient and the gauge parameter is understood.

By splitting the covariant field in time and spatial components one can solve the double-trace constraint; one can choose e.g.\ to encode the independent components of the spin-$s$ field in the traceful spatial tensors $\vf_{i_1 \cdots i_s}$, $\vf_{0i_1 \cdots i_{s-1}}$, $\vf_{00i_1 \cdots i_{s-2}}$ and $\vf_{000i_1 \cdots i_{s-3}}$. Within this set one has to distinguish between canonical variables and Lagrange multipliers. As discussed in sect.~\ref{sec:spin3-gauge}, combinations whose variations \eqref{fronsdal-gauge} do not contain time derivatives are canonical variables having conjugate momenta such that the Legendre transformation is invertible. These are the spatial components $\vf_{i_1 \cdots i_s}$ of the field and
\be
\a_{i_1 \cdots i_{s-3}} \equiv f^{-3} \vf_{000i_1 \cdots i_{s-3}} - 3 f^{-1} g^{kl} \vf_{0i_1 \cdots i_{s-3}kl} \, ,
\ee
where $f$ denotes the lapse in a static parametrisation of the AdS metric as in \eqref{defN}. The remaining independent components of the covariant field are Lagrange multipliers (see their gauge transformations in \eqref{var-multipliers}). We denote them as
\be
\cN_{i_1 \cdots i_{s-2}} \equiv f^{-1} \vf_{00i_1 \cdots i_{s-2}} \, , \qquad N_{i_1 \cdots i_{s-1}} \equiv \vf_{0i_1 \cdots i_{s-1}} \, .
\ee
As showed in \cite{Hbose}, one can then rewrite the action \eqref{fronsdal-s} in the form
\be \label{action-s}
S = \int\! d^{\,d}x\, \Big\{ \P\,\dot{\vf} + \tilde{\P}\,\dot{\a} - \cH - \cN\,\cC_{s-2} - N\,\cC_{s-1} \Big\} \, ,
\ee
where $\cH$ and the tensorial densities $\cC_{s-1}$ and $\cC_{s-2}$ only depend on the canonical variables and their conjugate momenta. The latter are related to the other fields by
\begin{align}
\P & = \frac{\sqrt{g}}{f} \sum_{n=0}^{\left[\frac{s}{2}\right]} \!\binom{s}{2n} g^n \Big\{ (1-2n) \dot{\vf}^{[n]} + (2n-1)(s-2n) \nabla N^{[n]} + 2n(2n-1) \nabla\!\cdot\! N^{[n-1]} \nn \\
& + \frac{n(s-2n)}{2}\, f \left( \nabla \a^{[n-1]} - \G \a^{[n-1]} \right) + n(n-1) f \left( \nabla\!\cdot \a^{[n-2]} - \G\cdot \a^{[n-2]} \right) \Big\} \, , \label{p1} \\[5pt]
\tilde{\P} & = \frac{\sqrt{g}}{f} \sum_{n=1}^{\left[\frac{s-1}{2}\right]} \frac{n}{2} \binom{s}{2n+1} g^{n-1} \Big\{ \dot{\a}^{[n-1]} + (s-2n-1) \nabla \cN^{[n]} + (2n+1) \nabla\!\cdot \cN^{[n-1]} \Big\} \, , \label{p2}
\end{align}
where we denoted by $\G$ the ``extrinsic'' Christoffel symbol $\G^0{}_{0i}$ defined in \eqref{christoffel}. For instance, $\G \a^{[n-1]}$ stands for the symmetrisation of the index carried by $\G^0{}_{0i}$ with the free indices in the $(n-1)$-th trace of the tensor $\a$.

The computation of surface charges does not require knowledge of the Hamiltonian $\cH$. For this reason we refer to \cite{Hbose} for an account of the Hamiltonian form of Fronsdal's action (in Poincar\'e coordinates) and we focus on the constraints, which are relevant for our analysis. As in the \mbox{spin-3} case, the constraints $\cC_{s-1} = 0$ and $\cC_{s-2} = 0$ are of first class and there are no secondary constraints \cite{Hbose}.
Consequently, they can be reconstructed from the gauge transformations they generate, i.e.\ from \eqref{fronsdal-gauge}. More specifically, one can compute
\be \label{generator}
\cG[\x ,\l] = \int d^{d-1}x \left( \x\, \cC_{s-1} + \l\, \cC_{s-2} \right) + Q_1[\x] + Q_2[\l]
\ee
by integrating the variations
\be \label{var-canonical}
\d \vf = \frac{\d \cG[\x,\l]}{\d\P} \, , \quad
\d \P = - \frac{\d \cG[\x,\l]}{\d\vf} \, , \quad 
\d \a = \frac{\d \cG[\x,\l]}{\d \tilde{\P}} \, , \quad
\d \tilde{\P} = - \frac{\d \cG[\x,\l]}{\d \a} \, ,
\ee
that one can derive from \eqref{fronsdal-gauge}. In the previous expressions $\x$ and $\l$ are tensors of rank $s-1$ and $s-2$, like the constraints with whom they are contracted.

Following this procedure, one obtains for the time-like spin-$s$ diffeomorphisms,
\begin{align}
& \cC_{s-2} = -\, s\, \nd \P - \sqrt{g} \sum_{n=0}^{\left[\frac{s-1}{2}\right]} \binom{s}{2n+1}\, g^n\, \bigg\{\, n\, \bigg[\, \frac{(2n+1) \triangle + m_n}{2}\, \a^{[n-1]} \nn \\
& + (n-1)(2n+1) \nd\nd \a^{[n-2]} + (s-2n-1)\frac{4n+1}{2}\, \nabla \nd \a^{[n-1]} + \binom{s-2n-1}{2} \nabla^2 \a^{[n]} \nn \\
& + (2n+1)\, \G_{k} \left( (n-1) \nd \a^{[n-2]\,k} + \frac{1}{2}\,\nabla^k \a^{[n-1]} \right) + (s-2n-1) \left( \frac{2n+1}{2}\, \G\, \nd \a^{[n-1]} \right. \nn \\
& \left. + (n+1) \G_{k} \nabla \a^{[n-1]\,k} - \frac{4n+3}{2}\, \G\, \G\cdot \a^{[n-1]} \right) - (2n+1)(n-1)\, \G_{k} \G_{l}\, \a^{[n-2]\,kl} \nn \\
& - \frac{2n+1}{2} \left( \G \cdot \G\right) \a^{[n-1]} \bigg] + (n+1) \binom{s-2n-1}{2} \left(\, \G \nabla \a^{[n]} - \G^2 \a^{[n-1]} \,\right) \bigg\} \, , \label{C(s-2)}
\end{align}
where $\triangle \equiv g^{ij} \pr_i \pr_j$ and the mass coefficients in the first line read
\be
\begin{split}
m_n = & - (2n+1) \left[ (s-2n-1)(d+s-2n-5) + (d+4s-11) - 2n(2n-1) \right] \\
& + 2(n+1)(s-2n-1) \, .
\end{split}
\ee
In several contributions we displayed some indices in order to avoid ambiguities as e.g.\ in
\be \label{example}
g^n\, \G_{k} \nabla \a^{[n-1]\,k} \equiv \underbrace{g_{ii} \cdots g_{ii}}_{\textrm{$n$ times}}\, \G^0{}_{0k} \nabla_{\!i\,} \a^k{}_{i_{s-2n-2}\,l_{2n-2}} \underbrace{g^{ll} \cdots g^{ll}}_{\textrm{$n-1$ times}} \, .
\ee
On the right-hand side of \eqref{example}, we also denoted symmetrisations by repeated covariant or contravariant indices, while the indices carried by a tensor are denoted by a single label with a subscript indicating their total number.

The constraint which generalises the generator of spatial diffeomorphisms reads instead
\begin{align}
& \cC_{s-1} = 3\, \nabla \tilde{\P} + (s-3)\, g\, \nd \tilde{\P} - \sqrt{g} \sum_{n=1}^{\left[\frac{s}{2}\right]} n \binom{s}{2n} g^{n-1} \bigg\{ \left(\, n \triangle - m'_n \right) \vf^{[n]} \nn \\
& + (n-2)(2n-1) \nd\nd \vf^{[n-1]}  + \frac{(4n-3)(s-2n)}{2}\, \nabla \nd \vf^{[n]} + \binom{s-2n}{2} \nabla^2 \vf^{[n+1]} \nn \\
& - \frac{(2n+1)(s-2n)}{2}\, \G_{k} \nabla \vf^{[n]\,k} - \frac{s-2n}{2}\, \G \left[\, (s-2n-1) \nabla \vf^{[n+1]} + 2(n-1) \nd \vf^{[n]} \,\right] \nn \\
& - (n-1)\, \G_{k} \left[\, (2n-1) \nd \vf^{[n-1]\,k} + \nabla^k \vf^{[n]} \,\right] + \binom{s-2n}{2} \G^2 \vf^{[n+1]} \nn \\
& + \frac{(4n-1)(s-2n)}{2}\, \G\, \G\cdot \vf^{[n]} + (n-1) \left[\, (2n-1)\, \G_k \G_l \vf^{[n-1]\,kl} + \left( \G \cdot \G \right)  \vf^{[n]} \,\right] \bigg\} \, , \label{C(s-1)}
\end{align}
where the mass coefficients in the first line are
\be
m'_n = 2n(s-2n)(D+s-3)-4n^2(s-5)+2n(3D+6s-26)-(4D+5s-22) \, .
\ee
The gauge variations of coordinates and momenta that we used to derive \eqref{C(s-2)} and \eqref{C(s-1)} are collected in Appendix~\ref{app:fronsdal-s}.

The boundary terms in \eqref{generator} that cancel the contributions from the integrations by parts needed to rewrite the variation of the generator as a bulk integral are
\begin{subequations} \label{Qgen}
\be \label{Q1gen}
\begin{split}
Q_1[\x] & = \int d^{d-2}S_k \bigg\{ s \, \x\, \P^{k} + \sqrt{g} \sum_{n=1}^{\left[\frac{s-1}{2}\right]}\! \frac{n}{2} \binom{s}{2n+1} \Big\{ (2n+1)\, \x^{[n]} \big[ \nabla^k \a^{[n-1]} \\
& + 2(n-1) \nd \a^{[n-2]\,k} \big] - (2n+1)\, \a^{[n-1]} \big[ \nabla^k \x^{[n]} + 2(n-1) \nd \x^{[n-1]\,k} \big] \\[6pt]
& + (s-2n-1)\, \x^{[n]\,k} \big[ (4n+1) \nd \a^{[n-1]} + (s-2n-2) \nabla \a^{[n]} \big] \\[5pt]
& - (s-2n-1)\, \a^{[n-1]\,k} \big[ (4n+1) \nd \x^{[n]} + (s-2n-2) \nabla \x^{[n+1]} \big] \\[3pt]
& + (2n+1) \Big[\, \x^{[n]} \big[ \G^k \a^{[n-1]} \! + 2(n-1)\, \G\cdot \a^{[n-2]\,k} \big] + 2n\, \a^{[n-1]} \G\cdot \x^{[n-1]\,k} \Big] \\
& + (s-2n-1)\, \big[ (2n+1)\, \x^{[n]}  \G\, \a^{[n-1]\,k} + 2(n+1)\, \a^{[n-1]}  \G\, \x^{[n]\,k} \big] \Big\} \bigg\} \, ,
\end{split}
\ee
\be \label{Q2gen}
\begin{split}
Q_2[\l] & = \int d^{d-2}S_k \bigg\{ - 3\, \l^k\, \tilde{\P} - (s-3) \l^{[1]} \tilde{\P}^{k}  + \sqrt{g} \sum_{n=0}^{\left[\frac{s}{2}\right]} n\binom{s}{2n} \Big\{ \l^{[n-1]}\! \left[\, n\, \nabla^k \vf^{[n]} \right. \\
& \left. +\, (n-2)(2n-1) \nd \vf^{[n-1]\,k} \,\right] + \frac{s-2n}{2}\, \l^{[n-1]\,k} \left[\, (s-2n-1) \nabla \vf^{[n+1]} \right. \\[4pt]
& \left. +\, (4n-3) \nd \vf^{[n]} \,\right] - \vf^{[n]} \left[\, n\, \nabla^k \l^{[n-1]} + (n-1)(2n-1) \nd \l^{[n-2]\,k} \,\right] \\[3pt]
& - \frac{s-2n}{2n}\, \vf^{[n]\,k} \left[\, (n-1)(s-2n-1) \nabla \l^{[n]} + n(4n-3) \nd \l^{[n-1]} \,\right] \\[4pt]
& -(n-1)\, \l^{[n-1]} \left[\, (s-2n) \G\, \vf^{[n]\,k} + (2n-1) \G\cdot \vf^{[n-1]\,k} + \G^k \vf^{[n]} \,\right] \\[3pt]
& - \frac{s-2n}{2}\, \l^{[n-1]\,k} \left[\, (s-2n-1) \G\, \vf^{[n+1]} + (2n+1) \G\cdot \vf^{[n]} \,\right] \Big\} \bigg\} \, .
\end{split}
\ee
\end{subequations}
Again, as for spin-$3$, to obtain \eqref{Qgen} we took advantage of the linear dependence on the fields of the variations $\delta Q_1$ and $\delta Q_2$, which implies their integrability. We also fixed the integration constants to zero on the zero solution.\footnote{The latter operation requires some care in $d = 3$, where it is customary to assign the negative energy $M = - 1/8G$ to AdS$_3$. To this end, one can either introduce an integration constant in \eqref{Qgen} for $s=2$ or declare that the vacuum is identified by a corresponding non-trivial $\cT_{IJ}$ in the boundary conditions specified below in \eqref{boundary-s}. This modification of the background metric does not affect the final expression for the charges \eqref{qfin}, since it introduces only corrections that vanish at the boundary. It however allows a direct match with the conventions of the Chern-Simons formulation \cite{HR,CS3} (see e.g.~\eqref{total-charge}).}
The resulting $Q_1$ and $Q_2$ are the surface charges: in the following we shall introduce boundary conditions on the canonical variables and on deformation parameters generating asymptotic symmetries. In sect.~\ref{sec:charges} we shall show how these conditions simplify asymptotically \eqref{Qgen} and we shall verify that the resulting spin-$s$ charges are finite and non-vanishing.

%%%%%%%%%%%%%%%%%%%%%%%%%%%%%%%%%%%%%%%%%%%%%%%%%%%%%%%%%%%%%%%%%%%%%
\subsection{Boundary conditions and asymptotic symmetries}\label{sec:bndH}
%%%%%%%%%%%%%%%%%%%%%%%%%%%%%%%%%%%%%%%%%%%%%%%%%%%%%%%%%%%%%%%%%%%%%

As in the spin-3 case, we derive boundary conditions on the canonical variables from the falloff at spatial infinity of the solutions of Fronsdal's equation in a convenient gauge, adopting the subleading branch. This is recalled in Appendix~\ref{app:boundary}: in the coordinates \eqref{poincare}, the relevant solutions behave at spatial infinity ($r \to \infty$) as
\begin{subequations} \label{boundary-s}
\begin{align} \label{boundaryspins}
\varphi_{I_{1}\cdots I_{s}} & = r^{3-d}\, \cT_{I_{1}\cdots I_{s}}(x^M) + \cO(r^{1-d}) \, , \\[5pt]
\varphi_{r\cdots rI_{1}\cdots I_{s-k}} & = \cO(r^{3-d-3k}) \, ,
\end{align}
\end{subequations}
where capital Latin indices denote again directions transverse to the radial one. The symmetric boundary tensor $\cT_{I_1 \cdots I_s}$ is a \emph{traceless conserved current}:
\be \label{cons-curr-s}
\pr^J \vf_{JI_1 \cdots I_{s-1}} = \h^{JK} \vf_{JKI_1\cdots I_{s-2}} = 0 \, .
\ee
For $s=2$ the fall-off conditions on $h_{IJ}$ and $h_{rI}$ agree with those proposed for non-linear Einstein gravity in eq.~(2.2) of \cite{AdS-generic}. Our $h_{rr}$ decays instead faster at infinity; the mismatch can be interpreted anyway as a radial gauge fixing, as discussed in sect.~2.1 of \cite{metric3D}. For $d=3$, \eqref{boundary-s} also agrees with the boundary conditions in eq.~(5.3) of \cite{metric3D}, with the same provisos as those discussed for the spin-3 case in sect.~\ref{sec:spin3-bnd}.

From \eqref{boundary-s} one obtains the boundary conditions on half of the canonical variables:
\begin{subequations}\label{boundary-spins-varphi}
\begin{align} 
\varphi_{\alpha_{1}\cdots \alpha_{s}} &= r^{3-d}\,{\cal{T}}_{\alpha_{1} \cdots \alpha_{s}}+{\cal{O}}(r^{1-d}) \, , \\[5pt]
\varphi_{r\cdots r\alpha_{1}\cdots \alpha_{s-k}} &= {\cal{O}}(r^{3-d-3k}) \, , \\[5pt]
\alpha_{\alpha_{1}\cdots \alpha_{s-3}} &= -2\,r^{-d}\,{\cal{T}}_{000\alpha_{1}\cdots \alpha_{s-3}}+{\cal{O}}(r^{-d-2}) \, , \\[5pt]
\alpha_{r\cdots r\alpha_{1}\cdots \alpha_{s-3-k}} &={\cal{O}}(r^{-d-3k}) \, ,
\end{align}
\end{subequations}
where Greek indices from the beginning of the alphabet denote angular coordinates on the $d-2$ sphere. The boundary conditions on the momenta are determined by making use of \eqref{boundary-s} in the definitions \eqref{p1} and \eqref{p2}:
\begin{subequations}\label{boundary-spins-varP}
\begin{align} \label{boundary-spins-Pi}
\Pi^{r\cdots r\alpha_{1} \cdots \alpha_{s-k}} &= \left\lbrace
\begin{array}{ll}
{\cal{O}}(r^{-1-2(s-k)})\, , \ & k=2m-2 \, , \\[5pt]
{\cal{O}}(r^{-2(s-k)})\, , \ & k=2m-1 \, ,
\end{array}
\right. \\[15pt]
\label{boundary-spins-P}
\tilde{\Pi}^{r\cdots r \alpha_{1} \cdots \alpha_{s-3-k}} &= \left\lbrace
\begin{array}{ll}
{\cal{O}}(r^{2-2(s-k)})\, , \quad & k=2m-2 \, , \\[5pt]
{\cal{O}}(r^{3-2(s-k)})\, , \quad & k=2m-1 \, ,
\end{array}
\right.
\end{align}
\end{subequations}
with $m$ a positive integer. For arbitrary spin, the components actually contributing to the charges are only $\P^{r\a_1\cdots\a_{s-1}}$ and $\tilde{\P}^{r\a_1\cdots\a_{s-4}}$, whose dependence on the boundary current $\cT_{I_1\cdots I_s}$ is detailed, respectively, in \eqref{Pr} and \eqref{Ptr}.

The deformation parameters that generate, via \eqref{generator} and \eqref{var-canonical}, gauge transformations preserving the boundary conditions behave at spatial infinity as
\begin{subequations}\label{bdn-deform-spins}
\begin{alignat}{5}
& \xi^{\alpha_{1}\cdots\alpha_{s-1}}=\chi^{\alpha_{1}\cdots\alpha_{s-1}}+{\cal{O}}(r^{-2}) \,, \qquad 
& \xi^{r\cdots r\alpha_{1}\cdots\alpha_{s-1-k}} & =  {\cal{O}}(r^{k}) \,, \\[10pt]
& \lambda^{\alpha_{1}\cdots\alpha_{s-2}}=-2\,r\,\chi^{0\alpha_{1}\cdots\alpha_{s-2}}r+{\cal{O}}(r^{-1}) \,, \qquad
& \lambda^{r\cdots r\alpha_{1}\cdots\alpha_{s-2-k}} & =  {\cal{O}}(r^{1+k}) \, ,
\end{alignat}
\end{subequations}
where $\chi^{I_1 \cdots I_{s-1}}$ is a \emph{traceless conformal Killing tensor} for the Minkowski metric $\h_{IJ}$ \cite{algebra}, satisfying
\be \label{conf-kill}
\pr_{(I_1} \chi_{I_2 \cdots I_{s})} - \frac{s-1}{d+2(s-2)-1}\, \h_{(I_1I_2\,} \prd \chi_{I_3 \cdots I_{s})} = 0 \, , \qquad \h_{JK} \chi^{JKI_1 \cdots I_{s-3}} = 0 \, .
\ee
The conditions \eqref{bdn-deform-spins} and \eqref{conf-kill} can be derived as for the spin-3 case discussed in sect.~\ref{sec:spin3-symm}. We refrain from detailing the analysis of asymptotic symmetries also for arbitrary spin because only the information on the leading terms displayed in \eqref{bdn-deform-spins} is relevant for the computation of charges. The existence of non-trivial asymptotic symmetries --~in which part of the subleading terms in \eqref{bdn-deform-spins} are fixed in order to preserve \eqref{boundary-s}~-- is again guaranteed by the existence of traceless Killing tensors for the AdS background (see e.g.~\cite{killing,kill_tensors_1,HScharges}). The latter solve the equations
\be \label{kill-s}
\bar{\nabla}_{\!(\m_1} \Lambda_{\m_2 \cdots \m_s)} = 0 \, , \qquad\qquad \L^\r{}_{\r\m_1 \cdots \m_{s-3}} = 0 \, ,
\ee
and generate gauge transformations preserving the vacuum solution $\vf_{\m_1 \cdots\m_s} = 0$, which is trivially included in the boundary conditions \eqref{boundary-s}.

%%%%%%%%%%%%%%%%%%%%%%%%%%%%%%%%%%%%%%%%%%%%%%%%%%%%%%%%%%%%%%%%%%%%%
\subsection{Charges}\label{sec:charges}
%%%%%%%%%%%%%%%%%%%%%%%%%%%%%%%%%%%%%%%%%%%%%%%%%%%%%%%%%%%%%%%%%%%%%

The computation of asymptotic charges for arbitrary integer spin $s$ is performed following similar steps to those for the spin-3 case, although with the contribution of terms that vanish for $s=3$, in both charges and momenta. Using the coordinates \eqref{poincare} and the asymptotic behaviour of the canonical variables (\ref{boundary-spins-varphi})--(\ref{bdn-deform-spins}) one sees that the only terms that contribute to (\ref{Qgen}) at the boundary are
\begin{subequations} \label{qfin}
\be\label{q1}
\begin{split}
\lim\limits_{r\rightarrow\infty}Q_1[\xi] & = s \!\int\! d^{d-2}x \bigg\{\, \xi\, \Pi^{r} + \sqrt{g} \sum_{n=1}^{\left[\frac{s-1}{2}\right]} \frac{n}{2} \binom{s-1}{2n} \Big\{\, \xi^{[n]} \left[ \nabla^r \a^{[n-1]} - \G^r \a^{[n-1]} \right. \\
& \left. +\, 2(n-1) \nabla\cdot \a^{[n-2]\,r} \right] - \a^{[n-1]} \left[ \nabla^r \xi^{[n]} + 2(n-1) \nabla\cdot \xi^{[n-1]\,r} \right] \Big\} \bigg\}
\end{split}
\ee
and
\begin{align}
& \lim\limits_{r\rightarrow\infty}Q_2[\lambda] = \!\int\! d^{d-2}x\bigg\{\! - (s-3) \lambda^{[1]} \tilde{\Pi}^{r}\! + \sqrt{g} \sum_{n=0}^{\left[\frac{s}{2}\right]} n\binom{s}{2n} \Big\{ \lambda^{[n-1]}\! \left[ (n-2)(2n-1) \nabla\!\cdot \varphi^{[n-1]\,r} \right. \nn \\
& \left. +\, n\, \nabla^r \varphi^{[n]}\,\right] -  \varphi^{[n]} \left[\, n\, \nabla^r \lambda^{[n-1]} + (n-1)\left(\, (2n-1) \nabla\cdot \lambda^{[n-2]\,r} + \G^r \lambda^{[n-1]} \,\right) \right] \Big\}\bigg\} \, , \label{q2}
\end{align}
\end{subequations}
where, as in \eqref{Q-final}, all omitted indices (including those involved in traces) take values in the $d-2$ sphere at infinity. 

The explicit form of these contributions in terms of the boundary current and of the boundary conformal Killing tensor that generates asymptotic symmetries is (see Appendix~\ref{app:charges} for details)
\begin{subequations} \label{Qgen-chi}
\begin{align}
\lim\limits_{r\rightarrow\infty}Q_{1} & = C \!\int\! d^{d-2}x\sum_{n=0}^{[(s-1)/2]}\binom{s-1}{2n} (\chi^{[n]})^{\a_{s-2n-1}} ({\cal{T}}^{\ [n]})_{0\a_{s-2n-1}} \, , \label{Qgen-chi1} \\
\lim\limits_{r\rightarrow\infty}Q_{2} & = C \!\int\! d^{d-2}x\sum_{n=1}^{[s/2]}\binom{s-1}{2n-1} (\chi^{[n-1]})^{0\a_{s-2n}} ({\cal{T}}^{[n]})_{\a_{s-2n}} \, , \label{Qgen-chi2}
\end{align}
\end{subequations}
with $C=s(d+2s-5)$. For better clarity, we specified the number of angular indices omitted in \eqref{qfin}, and we recall that traces are understood to follow from contractions with the metric $g_{\a\b}$ on the $d-2$ sphere. Taking advantage of the trace constraints on both ${\cal{T}}_{I_{s}}$ and ${\cal{\chi}}^{I_{s-1}}$, the sum of the two charges partly covariantises in the indices transverse to the radial direction as in \eqref{charge-cov}:
\be \label{total-charge}
\begin{split}
Q & \equiv \lim\limits_{r\rightarrow\infty}(Q_{1}+Q_{2})=
C \!\int\! d^{d-2}x\sum_{n=0}^{s-1}\binom{s-1}{n}\chi^{\supsuboverbrac[n]{0 \cdots 0}\, \a_{s-n-1}} {\cal{T}}_{\supsuboverbrace[n+1]{0 \cdots 0}\,\a_{s-n-1}} \\
& = s(d+2s-5) \!\int\!d^{d-2}x\, \chi^{I_{s-1}}{\cal{T}}_{0I_{s-1}} \, .
\end{split}
\ee
The charge $Q$ is manifestly preserved by time evolution, since it is the integral of the time component of a current which is conserved thanks to \eqref{cons-curr-s} and \eqref{conf-kill}. The final rewriting of $Q$ also manifests a chiral splitting in three space-time dimensions as in \eqref{3Dcharge}.

%%%%%%%%%%%%%%%%%%%%%%%%%%%%%%%%%%%%%%%%%%%%%%%%%%%%%%%%%%%%%%%%%%%%%
\section{Conclusions}\label{sec:conclusions}
%%%%%%%%%%%%%%%%%%%%%%%%%%%%%%%%%%%%%%%%%%%%%%%%%%%%%%%%%%%%%%%%%%%%%

We identified surface charges in AdS Fronsdal's theory as the boundary terms which enter the canonical generator of gauge transformations. This gives the charges \eqref{spin3-charges} in the \mbox{spin-3} example and the charges \eqref{Qgen} in the \mbox{spin-$s$} case. As discussed at the end of sect.~\ref{sec:spin3-charges}, these results are the analogues of the higher-spin charges identified by covariant methods in \cite{HScharges}. They are conserved when evaluated on-shell and on deformation parameters which generate gauge transformations leaving the AdS background strictly invariant. Both conditions can be however weakened: the field equations need to be fulfilled only asymptotically through suitable boundary conditions on the canonical variables and the residual gauge transformations need only to preserve these boundary conditions, without being exact Killing tensors everywhere.\footnote{Alternatively one can fix the gauge everywhere and not only asymptotically, and classify the gauge transformations preserving it; see sect.~2 of \cite{superrotations} for more comments on the relation between asymptotic conditions and gauge fixings.} This is the typical setup within the canonical approach we employed \cite{Regge,benguria-cordero}, and it allows one to discover infinitely many conserved charges in three space-time dimensions.  We specified boundary conditions in sects.~\ref{sec:spin3-bnd} and \ref{sec:bndH} and the resulting, greatly simplified, asymptotic expression for the charges are collected in \eqref{Q-final} for the \mbox{spin-3} example and in \eqref{qfin} for the \mbox{spin-$s$} case. We also showed that, with our choice of boundary conditions, the asymptotic charges can be rearranged so as to result from the integral of the time component of a conserved current, built from the contraction of a boundary conserved current with a conformal Killing tensor (see \eqref{Q-chi} and \eqref{Qgen-chi}). This form of our outcome allows a direct contact with the charges associated to the global symmetries of the boundary theory in AdS/CFT scenarios (compare e.g.\ with \cite{review-Giombi,global-charges-vasiliev}).

Although we worked in the linearised theory, the fact that the conservation of the asymptotic charges \eqref{Q-final} or \eqref{qfin} only relies on the boundary conditions \eqref{boundary} or \eqref{boundary-s} suggests that both charges and boundary conditions may remain valid when switching on interactions, at least in certain regimes, -- the idea being that asymptotically the fields become weak and the linearised theory applies. This expectation is supported by notable examples: our charges and fall-off conditions coincide with those obtained within the canonical formulation of gravity \cite{AD1,HT,BrownHenneaux,AdS-generic} and within the Chern-Simons formulation of higher-spin gauge theories in three space-time dimensions \cite{HR,CS3,metric3D}. Color charges in Yang-Mills can be obtained through a similar procedure \cite{AD2}. See also e.g.~\cite{covariant-charges1} for more examples of models in which surface charges are linear in the fields. 

In spite of these reassuring concurrencies, one should keep two important facts in mind. 
\begin{itemize}
\item For AdS Einstein gravity coupled to scalar matter, the scalar field might have a back-reaction on the metric that is sufficiently strong to force a weakening of the boundary conditions \cite{scalar1,scalar2,scalar3,Henneaux:2002wm}.  This observation is relevant in the present context because all known interacting higher-spin theories in more than three space-time dimensions have a scalar field in their spectra. One can thus foresee both additional non-linear contributions to the charges and relaxed boundary conditions when considering backgrounds with non-trivial expectation values for scalar fields. For instance, the boundary conditions of the scalar greatly influence the nature of the boundary dual within the AdS/CFT correspondence \cite{review-Giombi,ABJ-HS}.  As the pure-gravity analysis played an important role in paving the way to the study of scalar couplings, we expect nevertheless that our results will be important to attack the more general analysis of higher-spin interacting theories, e.g.\ in a perturbative approach. We defer the study of these interesting questions to future work. 
\item Second, we focused on the subleading branch of the boundary conditions.  How to include the other branch, or a mixture of the two,  has been investigated in the Hamiltonian asymptotic analysis for scalar fields, where nonlinearities may arise \cite{scalar1,scalar2,scalar3}.   This question is of course of definite importance  for higher spin holography \cite{boundary-action,unfolding-holography,alternate-bnd-cond}.
\end{itemize}

Even when the surface integrals are determined by the linear theory without nonlinear corrections, the interactions play a crucial role  in the study of the asymptotic symmetry algebra. This can be seen in many ways.  For instance, the Poisson bracket of charges, that is their algebra, can be derived from their variation under gauge transformations preserving the boundary conditions. However, as shown in sects.~\ref{sec:spin3-symm} and \ref{sec:bndH}, in the linearised theory the charges are left invariant by asymptotic symmetries in all space-time dimensions except $d = 3$ (but there the variation only gives rise to a central term in the algebra). The fact that one must include interactions in order to derive the asymptotic symmetry algebra is not a higher-spin feature, though.  In general relativity, although the surface charges are correctly reproduced by the linear theory,  one has to know how the metric transforms under diffeomorphisms (and not only under their linearised version) in order to obtain the algebra of asymptotic symmetries (see e.g.~\cite{BrownHenneaux,Brown:1986ed}). A similar phenomenon arises for color charges in Yang-Mills theory.

This mechanism is also clearly displayed by the comparison between sects.~\ref{sec:spin3-symm} and \ref{sec:bndH} and the similar analysis performed in $d=3$ in \cite{metric3D}, where the effect of gravitational couplings and of the self-interactions of higher-spin fields manifests itself in the non-linear variations of the linear charges. Similarly, higher-spin Lie algebras emerge when considering commutators of gauge transformations induced by cubic vertices and restricted to gauge parameters that at linearised level leave the AdS background invariant \cite{cubic-symmetries-1,cubic-symmetries-2}. It will be interesting to compare this analysis with that of surface charges developed in this and the related papers \cite{HScharges,charges-fermions}. Note that the analysis of asymptotic symmetries in \cite{ABJ-HS,review-strings} also involves ingredients that go beyond the linearised regime, like the full knowledge of the higher-spin gauge algebra on which the bulk theory is built upon.   

Let us finally point out other possible extensions.  Our results apply to any space-time dimension but they do not cover all possible bosonic higher-spin gauge theories. When $d > 4$, mixed symmetry fields (see e.g.~\cite{review-mixed} for a review) are a new possibility. In AdS they loose almost all gauge symmetries they display in flat space, but the remaining ones should give rise to conserved charges generalising those discussed in this paper. An analysis along the lines we followed for symmetric fields may start from the actions of \cite{maxwell-like}, which constitutes at present the closest generalisation of the Fronsdal action known in closed form. Partially massless fields \cite{pm1,pm2,pm3} are other gauge fields which can be defined on constant curvature backgrounds and have less gauge symmetry than Fronsdal's fields.  This generalisation comprises gauge fields that do not propagate unitarily on AdS. It would be interesting to investigate whether one can define in this case boundary conditions allowing for non-trivial residual symmetries and supporting consequently non-trivial conserved charges. How such charges, if they exist, would fit with the classical instability of the partially massless theories would also deserve to be understood.

%%%%%%%%%%%%%%%%%%%%%%%%%%%%%%%%%%%%%%%%%%%%%%%%%%%%%%%%%%%%%%%%%%%%%
\subsection*{Acknowledgements}
%%%%%%%%%%%%%%%%%%%%%%%%%%%%%%%%%%%%%%%%%%%%%%%%%%%%%%%%%%%%%%%%%%%%%

We are grateful to G.~Barnich, N.~Boulanger, D.~Francia, A.~Perez, M.~Taronna and R.~Troncoso for useful discussions. A.C. and M.H. thanks the Munich Institute of Astro- and Particle Physics (MIAPP) for hospitality while the manuscript was in preparation. This work was partially supported by the ERC Advanced Grants ``SyDuGraM" and ``High-Spin-Grav", by FNRS-Belgium (convention FRFC PDR T.1025.14 and  convention IISN 4.4503.15) and by the ``Communaut\'e Fran\c{c}aise de Belgique" through the ARC program. 

%%%%%%%%%%%%%%%%%%%%%%%%%%%%%%%%%%%%%%%%%%%%%%%%%%%%%%%%%%%%%%%%%%%%%
%%%%%%%%%%%%%%%%%%%%%%%%%%%%%%%%%%%%%%%%%%%%%%%%%%%%%%%%%%%%%%%%%%%%%

\begin{appendix}

%%%%%%%%%%%%%%%%%%%%%%%%%%%%%%%%%%%%%%%%%%%%%%%%%%%%%%%%%%%%%%%%%%%%%
\section{Notation and conventions}\label{app:conventions}
%%%%%%%%%%%%%%%%%%%%%%%%%%%%%%%%%%%%%%%%%%%%%%%%%%%%%%%%%%%%%%%%%%%%%

We adopt the mostly-plus convention for the space-time metric and we denote by $d$ the dimension of space-time. The AdS radius $L$ is defined as
\be \label{commutator}
[ \bar{\nabla}_{\!\m} \,, \bar{\nabla}_{\!\n}] V_\r = \frac{1}{L^2}\! \left( g_{\n\r} V_\m - g_{\m\r} V_\n \right) .
\ee
In the static coordinates \eqref{AdS} the spatial metric satisfies \eqref{commutator} as the full space-time metric provided one substitutes everywhere $\m,\n \to i,j$.

We distinguish between four types of indices, depending on whether the time and/or radial coordinates are included or not. Greek letters from the middle of the alphabet include all coordinates, small Latin letters include all coordinates except $t$, capital Latin letters include all coordinates except $r$, while Greek letters from the beginning of the alphabet denote angular coordinates on the unit $d-2$ sphere. In summary:
\begin{alignat}{3}
\m,\n,\ldots & \in \{t,r,\phi^1,\ldots,\phi^{d-2}\} \, , \qquad
& i,j,\ldots & \in \{r,\phi^1,\ldots,\phi^{d-2}\} \, , \nn \\[3pt]
I,J,\ldots & \in \{t,\phi^1,\ldots,\phi^{d-2}\} \, , \qquad
& \a,\b,\ldots & \in \{\phi^1,\ldots,\phi^{d-2}\} \, . \label{conventions-indices}
\end{alignat}

Indices between parentheses (or square brackets) are meant to be (anti)symmetrised with weight one, i.e.\ one divides the (anti)symmetrised expression by the number of terms that appear in it, so that the (anti)symmetrisation of a (anti)symmetric tensor is a projection.

In sect.~\ref{sec:spin3} omitted indices denote a trace, whose precise meaning depends on the context: in covariant expressions a contraction with the full space-time metric $g_{\m\n}$ is understood, while contractions with the spatial metric $g_{ij}$ or with the boundary metric $g_{IJ}$ are understood in formulae displaying the corresponding types of indices.

In most of sect.~\ref{sec:spins} we omit instead all indices: in linear expressions we elide all free indices, which are meant to be symmetrised, while in expressions quadratic in the fields omitted indices are meant to be contracted. In order to avoid ambiguities, however, in some terms we display explicitly a pair of contracted indices, continuing to elide the remaining ones. Traces, symmetrised gradients and divergences are denoted by
\be \label{notation}
\vf^{[n]} \equiv \vf_{\m_1 \cdots \m_{s-2n}\l_1 \cdots \l_n}{}^{\!\l_1 \cdots \l_n} \, , \qquad
\bar{\nabla} \vf \equiv \bar{\nabla}_{\!(\m_1} \vf_{\m_2 \cdots \m_{s+1})} \, , \qquad
\bar{\nabla}\cdot\vf \equiv \bar{\nabla}^\l \vf_{\l\m_1 \cdots \m_{s-1}} \, . 
\ee
We recalled here covariant expressions as an example, but contractions may also run only over spatial or transverse indices depending on the context. Similar shortcuts are adopted to denote symmetrisations or contractions with the ``extrinsic'' Christoffel symbol \eqref{christoffel}:
\be
\G \vf \equiv \G^0{}_{0(i_1} \vf_{i_2 \cdots i_{s+1})} \, , \qquad
\G\cdot \vf \equiv g^{kl} \G^0{}_{0k} \vf_{li_1 \cdots i_{s-1}} \, .
\ee

In Appendix~\ref{app:boundary} and in some expressions of sect.~\ref{sec:spins} we reinstate indices with the following convention: repeated covariant or contravariant indices denote a symmetrisation, while a couple of identical covariant and contravariant indices denotes as usual a contraction. Moreover, the indices carried by a tensor are substituted by a single label with a subscript indicating their total number. For instance, the combinations in \eqref{notation} may also be presented as
\be \label{example-repeated}
\vf^{[n]} = \vf_{\m_{s-2n}} \, , \qquad
\bar{\nabla} \vf = \bar{\nabla}_{\!\m} \vf_{\m_s} \, , \qquad
\bar{\nabla}\cdot\vf \equiv \bar{\nabla}^\l \vf_{\l\m_{s-1}} \, . 
\ee

Finally, our Hamiltonian conventions are $\{q,p\} = 1$ and $\dot{F} = \{F, \cH\}$.

%%%%%%%%%%%%%%%%%%%%%%%%%%%%%%%%%%%%%%%%%%%%%%%%%%%%%%%%%%%%%%%%%%%%
\section{Hamiltonian form of Fronsdal's action}\label{app:fronsdal}
%%%%%%%%%%%%%%%%%%%%%%%%%%%%%%%%%%%%%%%%%%%%%%%%%%%%%%%%%%%%%%%%%%%%

This appendix collects some additional details on the rewriting of spin-$3$ Fronsdal's action in Hamiltonian form and on the derivation of first class constraints in the spin-$s$ case.

\subsection{Spin 3} \label{app:fronsdal-3}

Integrating by parts the time derivative, one can eliminate on any static background all terms with $\dot{\cN}_i$ and $\dot{N}_{ij}$ which appear in the expansion in components of the Fronsdal action \eqref{fronsdal-action}. This is because the lapse does not depend on time, so that this step of the computation works as in flat space. Eliminating time derivatives from $\cN_i$ and $N_{ij}$, however, does not suffice to identify them as Lagrange multipliers: one must also verify that they enter linearly the action. The cancellation of the quadratic terms requires instead that the metric \eqref{AdS} be of constant curvature since in general
\be \label{S-generic}
\begin{split}
S & = S_{\textrm{Fronsdal}} + \int\! d^{\,d}x\, \frac{3\sqrt{g}}{2f}\, \bigg\{ f \left( R_{0ij}{}^0 - \frac{1}{L^2}\, g_{ij} \right)\! \left( 2 \cN_k \vf^{ijk} - \cN^i \vf^j - g^{ij} N \a \right) \\
& + 4 \left[ N^{ij} [ \nabla_{\!k} \,, \nabla_{\!i} ] N_j{}^k  + \frac{d}{L^2}\! \left( N_{ij}N^{ij} + N^2 \right) - R_{0ij}{}^0\! \left( N^i{}_k N^{jk} + 2 N^{ij} N + g^{ij} N^2 \right) \right] \\
& - \left[ \cN^i [ \nabla_{\!j} \,, \nabla_{\!i} ] \cN^j + \frac{2d}{L^2}\, \cN_i\cN^i - R_{0ij}{}^0\! \left( 3\cN^i\cN^j + g^{ij}\cN_i\cN^i \right) \right] \!\bigg\} \, ,
\end{split}
\ee
where $S_{\textrm{Fronsdal}}$ is the same action as \eqref{action}, with the same $\cH$, $\cC_i$ and $\cC_{ij}$, while $R_{\m\n\r}{}^\s$ denotes the Riemann tensor of the background. 
In obtaining this result we used only
\be
\G^0{}_{00} = \G^0{}_{ij} = \G^i{}_{0j} = 0 \, , \qquad \G^i{}_{00} = f^2 g^{ij} \G^0{}_{0j}
\ee
together with \eqref{christoffel}, which hold for a generic static metric.

The first line in \eqref{S-generic} would provide an additional contribution to the constraints, that vanishes on a constant curvature background on account of
\be \label{const-curvature}
R_{\m\n\r\s} = - \frac{1}{L^2} \left( g_{\m\r} g_{\n\s} - g_{\m\s} g_{\n\r} \right) .
\ee
The second and third lines must instead vanish and this requires \eqref{const-curvature}. The missing cancellation of the terms quadratic in $N_i$ and $N_{ij}$ is the counterpart of the loss of the gauge symmetry \eqref{cov-gauge} on arbitrary backgrounds. Correspondingly, in the canonical formalism one  obtains first class constraints only on constant curvature backgrounds. 

\subsection{Arbitrary spin} \label{app:fronsdal-s}

In order to reconstruct the canonical generator of gauge transformations \eqref{generator} by integrating \eqref{var-canonical}, one has to reconstruct the gauge transformations of the canonical variables from Fronsdal's covariant gauge transformation \eqref{fronsdal-gauge}. To this end, one has to set a map between the components of the covariant gauge parameter $\L_{\m_1 \cdots \m_{s-1}}$ and the deformation parameters $\x_{i_1 \cdots i_{s-1}}$ and $\l_{i_1 \cdots i_{s-2}}$ that enter \eqref{generator}. To obtain dimensions compatible with the action \eqref{action-s} and to agree with the spin-3 example, we choose
\be \label{comp-par}
\x_{i_1 \cdots i_{s-1}} = \L_{i_1 \cdots i_{s-1}} \, , \qquad \l_{i_1 \cdots i_{s-2}} = 2f^{-1} \L_{0\,i_1 \cdots i_{s-2}} \, .
\ee
The other components of the covariant gauge parameter are not independent from the ones above due to Fronsdal's constraint. 

Combining \eqref{fronsdal-gauge} with \eqref{comp-par} one obtains the following variations for the canonical variables and their traces:
\begin{subequations} \label{var-can-trace}
\begin{align}
\d \vf^{[n]} & = 2n\, \nd \x^{[n-1]} + (s-2n) \nabla \x^{[n]} \, , \label{dphi} \\[10pt]
\d \a^{[n]} & = -(2n+3) \nd \l^{[n]} - (s-2n-3) \nabla \l^{[n+1]} \, . \label{dalpha}
\end{align}
\end{subequations}
The gauge transformations of the Lagrange multipliers and their traces are obtained in a similar fashion:
\begin{subequations} \label{var-multipliers}
\begin{align}
\d N^{[n]} & = \dot{\x}^{[n]} + n f \left\{ \nd \l^{[n-1]} - \G\cdot \l^{[n-1]} \right\} + \frac{s-2n-1}{2} f \left\{ \nabla \l^{[n]} - \G\, \l^{[n]} \right\} , \label{dN1} \\[10pt]
\d \cN^{[n]} & = \dot{\l}^{[n]} + 2f \left\{ n\, \nd \x^{[n]} - (2n+1)\, \G\cdot \x^{[n]} \right\} + (s-2n-2)f \left\{ \nabla \x^{[n+1]} - 2\, \G\, \x^{[n+1]} \right\} . \label{dN2}
\end{align}
\end{subequations}
Substituting \eqref{var-can-trace} and \eqref{var-multipliers} in the definitions of the momenta \eqref{p1} and \eqref{p2} one obtains
\begin{align}
& \d \P = \sqrt{g} \sum_{n=0}^{\left[\frac{s}{2}\right]} \binom{s}{2n}\, g^n\, \bigg\{ (n-1)\binom{s-2n}{2} \nabla^2 \l^{[n]} + n\, \bigg[ (n-1)(2n-1) \nd\nd\l^{[n-2]} \nn \\[2pt]
& + \left( n\,\triangle - \frac{n(s-2n)(D+s-2n-3)+(2n-1)(D+2s-2n^2+4n-8)}{L^2} \right)\!\l^{[n-1]} \nn \\[2pt]
& + \frac{(4n-3)(s-2n)}{2}\, \nabla \nd \l^{[n-1]} + \frac{s-2n}{2}\, \G \left( (s-2n-1) \nabla \l^{[n]} + (2n+1) \nd \l^{[n-1]} \right) \nn \\[4pt]
& + (n-1)\, \G_{k} \left( (s-2n) \nabla \l^{[n-1]\,k} + (2n-1) \nd \l^{[n-2]\,k} + \nabla^k \l^{[n-1]} \right) \bigg] \bigg\} \, , \label{dp1}
\end{align}
\begin{align}
& \d\tilde{\P} = \sqrt{g} \sum_{n=1}^{\left[\frac{s-1}{2}\right]} \frac{n}{2} \binom{s}{2n+1}\, g^{n-1}\, \bigg\{\,  2(n-1)(2n+1) \nd\nd \x^{[n-1]} \nn \\[2pt]
& + (2n+1)\! \left[ \triangle - \frac{(s-2n-1)(D+s-2n-4) + 2(D+2s-2n^2+3n-7)}{L^2} \right] \x^{[n]} \nn \\[3pt]
& + (s-2n-1)\! \left[\, (4n+1) \nabla \nd \x^{[n]} + (s-2n-2) \nabla^2 \x^{[n+1]} \,\right] \nn \\[9pt]
& - (s-2n-1)\, \G \left[\, (s-2n-2) \nabla \x^{[n+1]} + 2(n+1) \nd \x^{[n]} \,\right] \nn \\[4pt]
& - (2n+1)\, \G_{k} \left[\, (s-2n-1) \nabla \x^{[n]\,k} + 2n\, \nd \x^{[n-1]\,k} + \nabla^k \x^{[n]} \,\right] \bigg\} \, . \label{dp2}
\end{align}
To derive \eqref{dp1} and \eqref{dp2} we used identities that are valid only on a constant-curvature space-time, as
\be \label{dGamma}
\nabla_{i} \G^0{}_{0j} + \G^0{}_{0i} \G^0{}_{0j} = \frac{1}{L^2}\, g_{ij}
\ee
which follows from \eqref{const-curvature}. We also commuted covariant derivatives using, for instance,
\be
[\, \nabla_{\!k} , \nabla \, ]\, \l^{[n-1]\,k} = - \frac{1}{L^2} \left\{ (D+s-2n-3) \l^{[n-1]} - (s-2n-1) g \l^{[n]} \right\} .
\ee
%

%%%%%%%%%%%%%%%%%%%%%%%%%%%%%%%%%%%%%%%%%%%%%%%%%%%%%%%%%%%%%%%%%%%%%
\section{Covariant boundary conditions}\label{app:boundary}
%%%%%%%%%%%%%%%%%%%%%%%%%%%%%%%%%%%%%%%%%%%%%%%%%%%%%%%%%%%%%%%%%%%%%

In this appendix we recall the falloff at the boundary of AdS$_d$ of the solutions of Fronsdal's equations of motion (see also e.g.~\cite{Mikhailov,metsaev-solutions,boundary-action}). To achieve this goal we partially fix the gauge freedom, and we also exhibit the fall-off conditions on the parameters of the residual gauge symmetry (which include the traceless Killing tensors of AdS$_d$).

We set the AdS radius to $L=1$ and we work in the Poincar\'e patch parameterised as
\be \label{poincare-z}
ds^2 = \frac{1}{z^2} \left( dz^2 + \h_{IJ} dx^I dx^J \right) .
\ee 
In these coordinates the spatial boundary is at $z \to 0$. All results can be easily translated in the coordinates \eqref{poincare} used in the main body of the text ($z = 1/r$), in which the boundary is at $r \to \infty$. We denote by capital Latin indices all directions transverse to the radial one (including time). 

%%%%%%%%%%%%%%%%%%%%%%%%%%%%%%%%%%%%%%%%%%%%%%%%%%%%%%%%%%%%%%%%%%%%
\subsection{Falloff of the solutions of the free equations of motion}\label{sec:eom}
%%%%%%%%%%%%%%%%%%%%%%%%%%%%%%%%%%%%%%%%%%%%%%%%%%%%%%%%%%%%%%%%%%%%

We wish to study the space of solutions of the Fronsdal equation in AdS$_d$ \cite{fronsdal-AdS,fronsdal-AdS-D} which, in the index-free notation of sect.~\ref{sec:spins}, reads
\be \label{fronsdal}
\Box \vf - s\, \bar{\nabla}\! \left( \bar{\nabla}\cdot \vf - \frac{s-1}{2}\, \bar{\nabla} \vf^{[1]} \right) - \left( s^2 + (d-6)s - 2(d-3) \right) \vf
- 2 \binom{s}{2} g\, \vf^{[1]} = 0 \, .
\ee
To this end it is convenient to partially fix the gauge freedom \eqref{fronsdal-gauge} by imposing $\bar{\nabla} \cdot \vf = 0$ and $ \vf^{[1]} = 0$.
This gauge is reachable on-shell,\footnote{In flat space one easily sees that one can reach the gauge $\pr \cdot \vf = \vf^{[1]} = 0$ only on shell: imposing $\vf^{[1]} = 0$ implies $\prd \L = 0$, but the divergence of the field, which now transforms as $\d \prd \vf = \Box \L$, is not divergenceless. In the gauge $\vf^{[1]} = 0$ the solutions of the equations of motion satisfy however $\Box \vf - s\, \pr\prd \vf = 0$; taking a divergence one obtains $\prd\prd \vf = 0$ allowing to reach the desired gauge.} as it is proven e.g.\ in sect.~3.1 of \cite{Mikhailov}. The previous statement amounts to say that the space of solutions (modulo gauge transformations) of \eqref{fronsdal} is equivalent to the space of solutions of the Fierz system
\begin{subequations} \label{fierz}
\begin{align}
& \!\!\left[ \Box - \left( s^2 + (d-6)s - 2(d-3) \right) \right] \vf = 0 \, , \label{fierz-eom} \\[3pt] 
& \bar{\nabla} \cdot \vf = 0 \, , \label{fierz-div} \\[1pt]
& \vf^{[1]} = 0 \, , \label{fierz-trace}
\end{align}
\end{subequations}
which also describes the propagation of a free massless particle of spin $s$ in the AdS$_d$ background \cite{fierz}. Actually the conditions \eqref{fierz-div} and \eqref{fierz-trace} do not fix completely the gauge: the Fierz system admits gauge transformations still of the form $\d \vf = s \bar{\nabla} \L$, but with gauge parameters constrained as
\begin{subequations} \label{parameter}
\begin{align}
& \!\!\left[\, \Box - (s-1)(d+s-3) \,\right] \L = 0 \, , \label{par-eom} \\[3pt]
& \bar{\nabla} \cdot \L = 0 \, , \label{par-div} \\[1pt]
& \L^{[1]} = 0 \, . \label{par-trace}
\end{align}
\end{subequations}

To analyse the falloff at the boundary of the solutions of the conditions \eqref{fierz}, one has to treat separately field components with a different number of indices along the $z$ direction. We denote them as
\be \label{radial-label}
\vf_{z_kI_{s-k}} \equiv \vf_{z \cdots z\,I_1 \cdots I_{s-k}} \, .
\ee
The divergence constraint \eqref{fierz-div} then gives
\be \label{divergence}
\left( z\pr_z - (d-2) \right) \vf_{z\m_{s-1}} + z\, \pr^I \vf_{I\m_{s-1}} = 0 \, ,
\ee
where Greek letters denote indices that take values in all space-time dimensions (including $z$).
The trace constraint \eqref{fierz-trace} gives instead
\be \label{trace}
\vf_{zz\m_{s-2}} + \h^{IJ} \vf_{IJ\m_{s-2}} = 0 \, .
\ee
Using \eqref{divergence} and \eqref{trace} the components of the equation of motion \eqref{fierz-eom} read
\be \label{eom}
\begin{split}
& \left[ z^2 \pr_z^2 - \left( d-2(s-k+1) \right) z\pr_z - (d+k-3)(2s-k-2) \right] \vf_{z_k I_{s-k}}\\[5pt]
& + z^2\, \Box \vf_{z_k I_{s-k}} - 2(s-k)\, z\,\pr_I \vf_{z_{k+1}I_{s-k-1}} + (s-k)(s-k-1)\, \h_{II} \vf_{z_{k+2}I_{s-k-2}} = 0 \, ,
\end{split}
\ee
where now $\Box = \h^{IJ} \pr_I\pr_J$ and repeated covariant or contravariant indices denote a symmetrisation, see e.g.~\eqref{example-repeated}.

Eq.~\eqref{divergence} implies $\vf_{z_k I_{s-k}} \sim z^{\D + k}$; substituting this ansatz in \eqref{eom} the terms in the second line are subleading for $z \to 0$ and the first line vanishes provided that
\be \label{fall-off-field}
\vf_{z\cdots z\, I_1 \cdots I_{s-k}} \sim z^{\D_\pm+\,k} \quad \textrm{with} \ \
\left\{
\begin{array}{l}
\D_+ = d-3 \\[5pt]
\D_- = 2-2s
\end{array}
\right. .
\ee
For $s = 0$ one recovers the asymptotic behaviour of the conformally coupled scalar of mass $m^2 = - 2(d-3)$ that enters the Vasiliev equations. For higher spins the subleading $\D_+$ branch gives the boundary conditions usually considered within the AdS/CFT correspondence \cite{review-Giombi,review-strings} and adopted in the text, while the $\D_-$ branch has been considered in a holographic setup in \cite{boundary-action,unfolding-holography,alternate-bnd-cond}.

%%%%%%%%%%%%%%%%%%%%%%%%%%%%%%%%%%%%%%%%%%%%%%%%%%%%%%%%%%%%%%%%%%%%
\subsection{Residual gauge symmetry}\label{app:gauge}
%%%%%%%%%%%%%%%%%%%%%%%%%%%%%%%%%%%%%%%%%%%%%%%%%%%%%%%%%%%%%%%%%%%%

The constraints \eqref{parameter} force the gauge parameters to have a precise fall off at the boundary, which can be determined as above. The divergence and trace constraints give again
\begin{subequations}
\begin{align}
& \left( z\pr_z - (d-2) \right) \L_{z\m_{s-2}} + z\, \pr^I \L_{I\m_{s-2}} = 0 \, , \label{par-div-open} \\[5pt]
& \L_{zz\m_{s-3}} + \h^{IJ} \L_{IJ\m_{s-3}} = 0 \, . \label{par-trace-open}
\end{align} 
\end{subequations}
Eq.~\eqref{par-div-open} implies $\L_{z_k I_{s-k-1}} \sim z^{\Theta + k}$ and using the relations above in \eqref{par-eom} one gets
\be \label{par-eom-open}
\begin{split}
& \left[ z^2 \pr_z^2 - \left( d-2(s-k) \right) z\pr_z - (d+k-1)(2s-k-2) + z^2 \Box \right] \L_{z_k I_{s-k-1}}\\[5pt]
& - 2(s-k-1)\, z\pr_I \L_{z_{k+1}I_{s-k-2}} + (s-k-1)(s-k-2)\, \h_{II} \L_{z_{k+2}I_{s-k-3}} = 0 \, .
\end{split}
\ee
This equation is identical to \eqref{eom}, apart from the shift $s \to s-1$ and a modification in the mass terms which implies that the first line vanishes asymptotically when
\be \label{fall-off-x}
\x_{z\cdots z\, I_1 \cdots I_{s-k-1}} \sim z^{\Theta_\pm+\,k} \quad \textrm{with} \ \
\left\{
\begin{array}{l}
\Theta_+ = d-1 \\[5pt]
\Theta_- = 2-2s
\end{array}
\right. .
\ee

We solved a second order equation and, as a result, we obtained two allowed asymptotic behaviours for the gauge parameters. On the contrary, the Killing equation \eqref{kill-s} is of first order and Killing tensors admit only a given boundary falloff. To fix it, notice that the AdS background is left invariant by the same set of gauge transformations in both Fronsdal's and Fierz's formulations: the gauge parameters are traceless in both setups and a traceless Killing tensor is also divergenceless thanks to
\be
g^{\a\a} \bar{\nabla}_{\!\a} \L_{\a\m_{s-2}} = - \frac{s-2}{2}\, g^{\a\a} \bar{\nabla}_{\!\m} \L_{\a\a\m_{s-2}} = 0 \, .
\ee
Traceless Killing tensors must therefore display one of the two boundary behaviours above. The Killing equation $\bar{\nabla}_{\!\m} \x_{\m_{s-1}} = 0$ branches in components as
\be \label{killing}
\left[ z\pr_z + (2s-k-2) \right] \L_{z_k I_{s-k-1}} = \frac{s-k-1}{k+1} \left[ -z\, \pr_I \L_{z_{k+1} I_{s-k-2}} + (s-k-2) \h_{II} \L_{z_{k+2} I_{s-k-3}} \right] \, .
\ee
The right-hand side is subleading, and one thus sees that Killing tensors belong to the $\Theta_-$ branch of \eqref{fall-off-x}.

%%%%%%%%%%%%%%%%%%%%%%%%%%%%%%%%%%%%%%%%%%%%%%%%%%%%%%%%%%%%%%%%%%%%
\subsection{Initial data at the boundary}\label{sec:leading}
%%%%%%%%%%%%%%%%%%%%%%%%%%%%%%%%%%%%%%%%%%%%%%%%%%%%%%%%%%%%%%%%%%%% 

In this subsection we recall the constraints on the initial data at the boundary imposed by divergence and trace constraints, and how the number of independent components is further reduced by the residual gauge symmetry. To this end we denote the leading contributions in the $\D_+$ branch by
\begin{subequations} \label{branch-vev}
\begin{align}
\vf_{I_s} & = z^{\D_+} \cT_{I_s}(x^J) + \cO\!\left(z^{\D_+ + 2}\right) , \label{vev} \\[5pt]
\vf_{z_k I_{s-k}} & = z^{\D_+ + k}\, t^{(k)}_{I_{s-k}}(x^J) + \cO\!\left(z^{\D_+ + k + 2}\right) , \quad 1\leq k \leq s \, ,
\end{align}
\end{subequations}
and the leading contributions in the $\D_-$ branch by
\begin{subequations} \label{branch-source}
\begin{align}
\vf_{I_s} & = z^{\D_-} \Phi_{I_s}(x^J) + \cO\!\left(z^{\D_- + 2}\right) , \label{source} \\[5pt]
\vf_{z_k I_{s-k}} & = z^{\D_- + k}\, \phi^{(k)}_{I_{s-k}}(x^J) + \cO\!\left(z^{\D_- + k + 2}\right) , \quad 1\leq k \leq s \, .
\end{align}
\end{subequations}
The subleading terms can be expressed in terms of the leading ones via the field equations (see e.g.~\cite{metsaev-solutions}).

The tensors $\cT_{I_s}$ and $\Phi_{I_s}$ are boundary fields of conformal dimensions, respectively, $\Delta_c = d + s - 3$ and $\Delta_s = 2 - s$. They thus correspond to the \emph{conserved currents} and \emph{shadow fields} of \cite{shadows}.\footnote{If one performs a dilatation $x^\m \to \l x^\m$ the tensor $\vf_{I_1 \cdots I_s}$ transforms as $\vf_{I'_s} = \l^{-s} \vf_{I_s}$, while on the right-hand side of \eqref{vev} or \eqref{source} one has $z'^{\D_\pm} = \l^{\D_\pm} z^{\D_\pm}$. As a result, $\cT$ and $\Phi$ must transform as $\cT_{I'_s} = \l^{- ( \Delta_+ + s )} \cT_{I_s}$ and $\Phi_{I'_s} = \l^{- ( \Delta_- + s )} \Phi_{I_s}$, from where one reads the conformal dimensions. To obtain a direct matching between the exponents $\D_\pm$ and the conformal dimensions, one can contract all indices with the (inverse) vielbein $e_\m{}^M= z^{-1}\, \d_\m{}^M$ as e.g.\ in \cite{boundary-action}. }
Note that $\D_c + \D_s = d -1$, i.e.\ the sum of conformal dimensions is equal to the dimension of the boundary. Therefore the coupling $\Phi^{I_s} \cT_{I_s}$ is conformally invariant. In the AdS/CFT jargon $\Phi_{I_s}$ is a \emph{source} and $\cT_{I_s}$ is the corresponding \emph{vev}.

The trace constraint \eqref{trace} then implies that all tensors on the right-hand side of \eqref{branch-vev} and \eqref{branch-source} are traceless, since the trace of $\vf_{z_k I_{s-k}}$ is subleading with respect to the traceless part. The divergence constraint \eqref{divergence} has instead different consequences in the two branches: in the $\D_+$ branch of vevs one obtains
\begin{subequations}
\begin{align}
& \prd \cT_{I_{s-1}} = 0 \, , \label{conservation} \\[5pt]
& t^{(k+1)}_{I_{s-k-1}} = - \frac{1}{k}\, \prd t^{(k)}_{I_{s-k-1}} \, , \quad 1 \leq k \leq s \, ,
\end{align}
\end{subequations}
while in the $\D_-$ branch of sources one obtains
\be
\phi^{(k+1)}_{I_{s-k-1}} = \frac{1}{d+2s-k-5}\, \prd \phi^{(k)}_{I_{s-k-1}} \, , \quad 0 \leq k \leq s \, ,
\ee
where $\phi^{(0)}_{I_s} \equiv \Phi_{I_s}$ and where divergences are meant to involve sums only over indices transverse to $z$. Eq.~\eqref{conservation} shows that, as expected from its conformal weight, $\cT$ is a conserved current. Up to this point $t^{(1)}$ is instead an arbitrary tensor (but we still have to consider the residual gauge symmetry), while all other tensors in \eqref{branch-vev} are not independent. In the other branch the only independent tensor is instead $\Phi$, whose number of independent components is the same as those of $\cT$ plus $t^{(1)}$. 

The residual gauge symmetry is generated by
\be \label{par+}
\L_{z_k I_{s-k-1}} = z^{\Th_+ + k}\, \x^{(k)}_{I_{s-k-1}}(x^J) + \cO\!\left(z^{\Th_+ + k + 2}\right) , 
\ee
or
\be \label{par-}
\L_{z_k I_{s-k-1}} = z^{\Th_- + k}\, \ve^{(k)}_{I_{s-k-1}}(x^J) + \cO\!\left(z^{\Th_- + k + 2}\right) .
\ee
In the following we shall often denote $\x^{(0)}$ by $\x$ and $\ve^{(0)}$ by $\ve$.
As for the fields, the trace constraint \eqref{par-trace-open} imposes that all tensors in \eqref{par+} and \eqref{par-} be traceless. For gauge parameters the divergence constraint \eqref{par-div-open} has instead the same form in both branches:
\begin{subequations}
\begin{align}
& \x^{(k+1)}_{I_{s-k-2}} = - \frac{1}{k+2}\, \prd \x^{(k)}_{I_{s-k-2}} \, , \quad 0 \leq k \leq s-1 \, , \\[5pt]
& \ve^{(k+1)}_{I_{s-k-2}} = \frac{1}{d+2s-k-5}\, \prd \ve^{(k)}_{I_{s-k-2}} \, , \quad 0 \leq k \leq s-1 \, . \label{div-ve}
\end{align}
\end{subequations}
This means that only the tensors $\x_{I_{s-1}}$ and $\ve_{I_{s-1}}$ are independent.

The components of the field transform as
\be \label{var-residual}
\d \vf_{z_k I_{s-k}} = k \left[ \pr_z + \frac{2s-k-1}{z} \right]\! \L_{z_{k-1}I_{s-k}} + (s-k)\! \left[ \pr_I \L_{z_k I_{s-k-1}}\! - \frac{s-k-1}{z}\, \h_{II} \L_{z_{k+1}I_{s-k-2}} \right] .
\ee
In the $\Th_+$ branch the first contribution is leading and gives
\be
\d \vf_{z_k I_{s-k}} = k (d+2s-3) z^{d-3+k} \x^{(k-1)}_{I_{s-k}} + \cO(z^{d-1+k}) \, .
\ee
These gauge transformations act naturally on vevs, where they induce
\be
\d \cT_{I_s} = 0 \, , \qquad
\d t^{(1)}_{I_{s-1}} = (d+2s-3) \L_{I_{s-1}} \, .
\ee
The tensor $t^{(1)}$ is thus a Stueckelberg field that can be eliminated using the residual symmetry generated by $\x$ and the solution is fully specified by the conserved current $\cT$.

In the $\Th_-$ branch the coefficient of the leading contribution vanishes, and one obtains
\be
\begin{split}
\d \vf_{z_k I_{s-k}} & = \frac{z^{2-2s+k}}{d+2s-5} \Big[\, k\, \Box \ve^{(k-1)}_{I_{s-k}} + (s-k)(d+2(s-k)-5)\, \pr_I \ve^{(k)}_{I_{s-k-1}} \\
& - (s-k)(s-k-1)(d+2s-k-5)\, \h_{II} \ve^{(k+1)}_{I_{s-k-2}} \,\Big] + \cO(z^{4-2s+k}) \, ,
\end{split}
\ee
where we used the second-order equation \eqref{par-eom-open} to express the subleading contributions in $\L_{z_{k-1}I_{s-k}}$ in terms of the leading ones. These gauge transformations act naturally on sources, where they induce
\be \label{var-source}
\d \Phi_{I_s} = s\, \pr_{I} \ve_{I_{s-1}} - \frac{s(s-1)}{(d-1)+2(s-2)}\, \h_{II} \prd\ve_{I_{s-2}}
\ee
and similar transformations on the $\phi^{(k)} = \phi^{(k)}(\Phi)$.  To obtain \eqref{var-source} we used \eqref{div-ve} to eliminate $\ve^{(1)}$. This gauge freedom reduces the number of independent components of $\Phi_{I_s}$ such that it becomes identical to that of the conserved current $\cT_{I_s}$. Moreover, the gauge transformations \eqref{var-source} leave invariant the coupling $\Phi^{I_s} \cT_{I_s}$.

To summarise, the solutions of the field equations are specified asymptotically either by a traceless and conserved current $\cT_{I_s}$ or by a traceless tensor $\Phi_{I_s}$ subjected to the gauge symmetry \eqref{var-source}. These tensors enter the components transverse to $z$ as in \eqref{vev} and \eqref{source}, while all other components of the spin-$s$ field are expressed in terms of them via the field equations (or set to zero by the residual gauge symmetry \eqref{var-residual}).

%%%%%%%%%%%%%%%%%%%%%%%%%%%%%%%%%%%%%%%%%%%%%%%%%%%%%%%%%%%%%%%%%%%%%
\section{Conformal Killing tensors}\label{app:identities}
%%%%%%%%%%%%%%%%%%%%%%%%%%%%%%%%%%%%%%%%%%%%%%%%%%%%%%%%%%%%%%%%%%%%%

In this appendix we recall the structure of the general solution of the conformal Killing equations \eqref{conf-kill} and we present it explicitly in the rank-2 case. We also prove the identities \eqref{kill1} and \eqref{kill2}, that we used in the analysis of asymptotic symmetries in the spin-3 example.

\subsection*{General solution of the rank-2 conformal Killing equation}

When $d-1 > 2$, the solutions of the conformal Killing tensor equation \eqref{conf-kill} are in one-to-one correspondence with rectangular traceless Young tableaux with two rows of \mbox{$s-1$} boxes in $d+1$ dimensions \cite{algebra}, that we denote by $\{s,s\}$. For instance, a generic conformal Killing vector, $v_I = a_I - \o_{I|J} x^J + \l\, x_I + b_{K}\! \left( 2\,x_Ix^K - \d_I{}^{K} x^2 \right)$,
can be cast in the form \mbox{$v^I = \Phi_{\und{M}} V^{\und{M}|\und{N}} \,\Psi^I{}_{\!\und{N}}$}, where underlined indices take value in the $(d+1)$-dimensional ambient space $\mathbb{R}^{2,d-1}$ and
\be
V^{\und{I}|\und{J}} = 
\left(\begin{array}{ccc}
0 & 2b^J & \l \\
-2b^I & \omega^{I|J} & -a^I \\
-\l & a^J & 0
\end{array}\right) , \quad
\Phi_{\und{I}} = 
\left(\begin{array}{c}
-x^2/2 \\ x_I \\ 1
\end{array}\right) , \quad
\Psi^{I}{}_{\!\und{J}} =
\left(\begin{array}{c}
-x^I \\ \d^I{}_{\!J} \\ 0
\end{array}\right) .
\ee
A similar characterisation of traceless conformal Killing tensors exists for any value of the rank \cite{algebra}. 
In particular, the pattern of tensors in $d-1$ dimensions that specify the solution follows from the decomposition of a traceless $\{s,s\}$ tensor in $d+1$ dimensions (using the branching rules discussed e.g.\ in \S~8.8.A of \cite{Barut}). In the rank-2 case one obtains
\be
\{2,2\}_{o(d+1)} = \left( \{2,2\} + 2\,\{2,1\} + 3\,\{2\} + \{1,1\} + 2\,\{1\} + 1 \right){}_{\!o(d-1)} \, .  
\ee
Correspondingly, the general solution of \eqref{kill-boundary} is
\begin{align}
& \chi_{IJ} = a_{IJ} + \left( b_{(I}x_{J)} - \frac{1}{d-1}\, \h_{IJ}\, b\cdot x\right) + \o_{IJ|K}\,x^K + \l \left( x_I x_J - \frac{1}{d-1}\, \h_{IJ}\,x^2 \right) \nn \\
& + \r_{K|(I}x_{J)}x^K + \left(  2\, c_{K(I} x_{J)}x^K  - c_{IJ}\, x^2 - \frac{2}{d-1}\, \h_{IJ} c_{KL} x^K x^L \right) + \O_{IJ|KL}\,x^K x^L \nn \\
& + \left( 2\,\tilde{b}_K x_Ix_Jx^K \! - \tilde{b}_{(I} x_{J)}x^2 - \frac{1}{d-1}\, \h_{IJ} (b\cdot x) x^2 \right) + \left( 2\,\tilde{\o}_{KL|(I}x_{J)}x^Kx^L \! + \tilde{\o}_{IJ|K}x^Kx^2 \right) \nn \\[5pt]
& + \tilde{c}_{KL} \left( 4\,x_Ix_Jx^Kx^L - 4\, \d^K{}_{\!(I} x_{J)} x^L x^2 + \d^K{}_{\!I} \d^L{}_{\!J}\, x^4  \right) . \label{sol-conf} 
\end{align}
All tensors in this expression are irreducible and traceless, so that e.g.\ $\o_{IJ|K}$ is symmetric in its first two indices and satisfies $\o_{(IJ|K)} = 0$.

When $d-1 = 2$, the general solution depends instead on two chiral functions for any value of the rank. Introducing the light-cone coordinates $x^\pm = t \pm \phi$, eqs.~\eqref{conf-kill} are solved by 
\be \label{sol-conf3}
\chi^{+ \cdots +} = \chi(x^+) \, , \qquad
\chi^{+\cdots+-\cdots-} = 0 \, , \qquad
\chi^{- \cdots -} = \tilde{\chi}(x^-) \, .
\ee

\subsection*{Proof of eqs.~\eqref{kill1} and \eqref{kill2}}

We wish to prove that the rank-2 conformal Killing equation \eqref{kill-boundary} implies the identities \eqref{kill1} and \eqref{kill2}, which entail the cancellation of the first two subleading orders in $\d \vf_{IJK}$. The following proof is independent on the space-time dimension, but when $d > 3$ one could also verify these identities by acting with the differential operators they involve on \eqref{sol-conf}.

As a first step, one can act with a derivative on the conformal Killing equation, and rewrite the result as follows
\be \label{1st-id}
\begin{split}
0 & = \pr_M \!\left( \pr_{(I} \chi_{JK)} - \frac{2}{d+1}\, \h_{(IJ\,} \prd \chi_{K)} \right) = 2 \left( \pr_{(M} \pr_I \chi_{JK)} - \frac{2}{d+1}\, \h_{(MI\,} \pr_J \prd \chi_{K)} \right) \\
& - \, \pr_{(I} \pr_J\chi_{K)M} - \frac{1}{d+1} \left(\, \h_{(IJ|} \pr_M \prd\chi_{|K)} - 2\, \h_{M(I} \pr_J \prd \chi_{K)} - \h_{(IJ} \pr_{K)} \prd \chi_M \right) .\end{split}
\ee
The first term on the right-hand side vanishes since it is the symmetrisation of the left-hand side. As a result, one discovers that the second line vanishes as well.

To prove \eqref{kill1} one needs another identity obtained in a similar fashion:
\begin{align}
0 & = 3\, \pr_M \pr_N \!\left( \pr_{(I} \chi_{JK)} - \frac{2}{d+1}\, \h_{(IJ\,} \prd \chi_{K)} \right) \nn \\
& = 10 \left( \pr_{(M} \pr_N \pr_I \chi_{JK)} - \frac{2}{d+1}\, \h_{(MN\,} \pr_I \pr_J \prd \chi_{K)} \right) - \pr_{I} \pr_J \pr_K \chi_{MN} \label{2nd-id} \\
& + \frac{2}{d+1}\, \Big\{ \h_{(IJ|} \pr_M \pr_N \prd\chi_{|K)} - 2 \left( \h_{M(I} \pr_{J|} \pr_N \prd \chi_{|K)} + \h_{N(I} \pr_{J|} \pr_M \prd \chi_{|K)} \right) \nn \\
& - \h_{(IJ} \pr_{K)} \pr_{(M} \prd \chi_{N)} + 2 \left( \h_{M(I} \pr_J \pr_{K)} \prd \chi_N + \h_{N(I} \pr_J \pr_{K)} \prd \chi_M \right) + \h_{MN} \pr_{(I} \pr_J \prd \chi_{K)} \Big\} . \nn
\end{align}
The first term on the right-hand side --~with a symmetrisation over five indices~-- vanishes again because it is the symmetrisation of the left-hand side. To reach this expression we also used the identity derived from \eqref{1st-id}.

One can finally contract the result with $\h^{MN}$ obtaining (recall that $\chi_{MN}$ is traceless): 
\be \label{3rd-id}
(d-1)\, \pr_{(I} \pr_J \prd \chi_{K)} - \h_{(IJ|} \!\left\{ \pr_{|K)} \prd\prd\prd \chi - \Box\, \prd\chi_{|K)} \right\} = 0 \, .
\ee
By computing two divergences of the conformal Killing equation \eqref{kill-boundary} one also obtains
\be
\Box\, \prd \chi_I = - \frac{d-3}{2d}\, \pr_{I\,} \prd\prd \chi \, ,
\ee
so that \eqref{3rd-id} implies \eqref{kill1} for $d>1$.

To prove the identity \eqref{kill2}, it is convenient to compute two gradients of \eqref{kill1} and to manipulate the resulting expression as in the previous subsection:
\begin{align}
0 & = 3\, \pr_M \pr_N \!\left( \pr_{(I} \pr_J \prd \chi_{K)} - \frac{3}{2d}\, \h_{(IJ} \pr_{K)} \prd\prd \chi \right) \nn \\
& = 5 \left( \pr_{(M} \pr_N \pr_I \pr_J \prd \chi_{K)} - \frac{3}{2d}\, \h_{(MN} \pr_I \pr_J \pr_{K)} \prd\prd \chi \right) - 2\, \pr_I \pr_J \pr_K \pr_{(M} \prd\chi_{N)} \\
& + \frac{3}{4d} \left\{ \h_{MN} \pr_I \pr_J \pr_K + 3 \left( \h_{M(I} \pr_J \pr_{K)} \pr_N + \h_{N(I} \pr_J \pr_{K)} \pr_M \right) - 3\, \h_{(IJ} \pr_{K)} \pr_M \pr_N \right\} \prd\prd\chi \, . \nn
\end{align}
The first term on the right-hand side vanishes. Contracting the remaining addenda with $\h^{MN}$ and taking into account \eqref{kill3} one obtains \eqref{kill2}.

%%%%%%%%%%%%%%%%%%%%%%%%%%%%%%%%%%%%%%%%%%%%%%%%%%%%%%%%%%%%%%%%%%%%%
\section{Spin-$s$ charges}\label{app:charges}
%%%%%%%%%%%%%%%%%%%%%%%%%%%%%%%%%%%%%%%%%%%%%%%%%%%%%%%%%%%%%%%%%%%%%

This appendix is dedicated to provide the reader with some details of the computation of the asymptotic charges in the general case of spin $s$. As mentioned in sect.~\ref{sec:spins}, taking into account the boundary conditions on canonical variables and deformation parameters (\ref{boundary-spins-varphi})--(\ref{bdn-deform-spins}) one sees that the finite contributions to the charges come from the terms
\be \label{breakdown-charges}
\lim\limits_{r\rightarrow\infty}Q_1[\xi] = \!\int\! d^{d-2}x\, \Big\{A_{1}(\Pi)+B_{1}(\alpha)\Big\} \, ,\quad
\lim\limits_{r\rightarrow\infty}Q_2[\lambda] =  \!\int\! d^{d-2}x\, \Big\{A_{2}(\tilde{\Pi})+B_{2}(\varphi) \Big\} \, ,
\ee
where
\begin{subequations}
{\allowdisplaybreaks
\begin{align}
A_{1}(\Pi)&\equiv s \, \xi \Pi^{r} \, , \label{A1} \\
B_{1}(\alpha)&\equiv s\,\sqrt{g} \sum_{n=1}^{\left[\frac{s-1}{2}\right]}\! \frac{n}{2} \binom{s-1}{2n} \Big\{ \xi^{[n]} \left[ \nabla^r \a^{[n-1]} + 2(n-1) \nabla\cdot \a^{r[n-2]} \right] \nn \\
& - \a^{[n-1]} \left[ \nabla^r \xi^{[n]} + 2(n-1) \nabla\cdot \xi^{r[n-1]} - \G^r \xi^{[n]} \right] \Big\} \, , \\[8pt]
A_{2}(\tilde{\Pi})&\equiv (3-s)\, \lambda^{[1]} \tilde{\Pi}^{r} \, , \\[1pt]
B_{2}(\varphi)&\equiv \sqrt{g} \sum_{n=0}^{\left[\frac{s}{2}\right]} n \binom{s}{2n} \Big\{ \lambda^{[n-1]} \left[\, n\, \nabla^r \varphi^{[n]} + (n-2)(2n-1) \nabla\cdot \varphi^{r[n-1]} \,\right] \nonumber\\
& -  \varphi^{[n]} \left[\, n\, \nabla^r \lambda^{[n-1]} + (n-1) \left( (2n-1) \nabla\cdot \lambda^{r[n-2]} + \G^r \lambda^{[n-1]} \right)\right] \Big\} \, .
\end{align}
}
\end{subequations}
At this stage, differently from \eqref{qfin}, the omitted indices in the expressions above still include all coordinates except time as in \eqref{Qgen}. Along the way we shall show that the contributions from radial components are actually subleading; eventually all omitted indices can thus be considered to be valued on the $d-2$ sphere at infinity as indicated in sect.~\ref{sec:charges}.

The contribution of the terms in $B_{1}$ and $B_{2}$ is computed in a straightforward fashion using the boundary conditions (\ref{boundary-spins-varphi}) and (\ref{bdn-deform-spins}). As an example, let us consider the first term in $B_{1}$, that is \mbox{$\sqrt{g}\, \xi^{[n]} \nabla^r \a^{[n-1]}$}. Displaying explicitly the free indices on each tensor as in Appendix~\ref{app:boundary}, one obtains
\be\label{Ck}
\sqrt{g}\,g^{rr}\xi^{i_{s-2n-1}}\nabla_{\!r}\alpha_{i_{s-2n-1}} = \sqrt{g}\,g^{rr} \sum_{k=0}^{s-2n-1} \binom{s-2n-1}{k}\, \xi^{r_k\a_{s-k-1}}\nabla_{\!r}\alpha_{r_k\a_{s-k-1}} \, .
\ee
We recall that small Latin indices include all coordinates except time, while Greek indices from the beginning of the alphabet do not include neither time nor radial directions. We also resorted to a collective notation for the radial indices as in \eqref{radial-label}.
Expanding the covariant derivative and using the boundary conditions one gets
\be
\nabla_{\!r}\alpha_{r_k\a_{s-k-1-2n}} \sim -2(3-d-k-s)\,r^{1-d-3k-2n} \, ,
\ee
and so
\be \label{contraction-example}
\sqrt{g}\,g^{rr}\xi^{r_k\a_{s-k-1-2n}}\nabla_{\!r}\alpha_{r_k\a_{s-k-1-2n}} \sim -2(3-d-k-s)\,r^{-2k} \, .
\ee
Clearly the only finite contribution when $r\to\infty$ comes from the term $k=0$ in (\ref{Ck}). This is a general feature that one also encounters in the analysis of the other contributions in $B_1$ (and more generally in \eqref{breakdown-charges}). One can compute them along similar lines and find
\be\label{B1}
B_{1}=C\sum_{n=1}^{[\frac{s-1}{2}]} \frac{n(s-1)!}{(2n)!(s-2n-1)!}\, \chi^{\supsuboverbrac[2n]{0\cdots0}\a_{s-2n-1}}{\cal{T}}_{0\supsuboverbrace[2n]{0\cdots0}\a_{s-2n-1}} \, ,
\ee
with $C\equiv s(d+2s-5)$.

Now let us turn to $A_{1}$: from \eqref{boundary-spins-varP} and \eqref{bdn-deform-spins} one sees that, similarly to e.g.~\eqref{contraction-example}, the only finite contribution in $\xi \Pi^{r}$ comes from the term $\xi_{\a_{s-1}} \Pi^{r\a_{s-1}}$. Its computation is however less direct since one has to take into account the definition \eqref{p1} of the momentum. To illustrate this point, consider the first term in $\Pi_{r \a_1\cdots \a_{s-1}}$:
\be \label{my-rewriting}
\begin{split}
\Pi_{r \a_{s-1}} & = \frac{\sqrt{g}}{f} \sum_{n=0}^{\left[\frac{s}{2}\right]} \binom{s}{2n}(2n-1)(s-2n)(g_{\a\a})^n \Big[\nabla_{\!r} N_{\a_{s-2n-1}} \\ & + (s-2n-1)\nabla_{\!\a} N_{r\a_{s-2n-2}} \big] D_{1}(n) + \cdots \, ,
\end{split}
\ee
where we have introduced the degeneracy factor $D_{1}(n)\equiv\frac{1}{(2n+1)!!}\binom{s}{2n+1}^{-1}$, which counts the number of equivalent terms after the symmetrisation over free indices.
Expanding the covariant derivative and taking advantage of the trace constraint in \eqref{conf-kill} one finds
\begin{subequations}
\begin{align}
\nabla_{\!r}N_{\a_{s-2n-1}} & \sim (4-d-s)\,r^{2-d-2n}\,\cT_{0\supsuboverbrace[2n]{0\cdots0}\,\a_{s-2n-1}} \, , \\
\nabla_{\!\a}N_{r\a_{s-2n-2}} & \sim -\,r^{2-d-2n}\,\cT_{0\supsuboverbrace[2n]{0\cdots0}\,\a_{s-2n-2}} \, ,
\end{align}
\end{subequations}
so that
\be
\Pi_{r \a_{s-1}} \sim \sum_{n=0}^{\left[\frac{s}{2}\right]} \binom{s}{2n}(2n-1)(s-2n) D_{1}(n)(5-d-2s+2n)\,r^{-2}\,(g_{\a\a})^n {\cal{T}}_{0\supsuboverbrace[2n]{0\cdots0}a_{s-2n-1}} \, .
\ee
Repeating the analysis for the other relevant terms in $\Pi_{r \a_1\cdots \a_{s-1}}$ one gets
\be \label{Pr}
\Pi_{r\a_{s-1}} \sim -\,s(d+2s-5)\, r^{-2}\sum_{n=0}^{[s/2]}D_{1}(n)\binom{s-1}{2n}(n-1)\,(g_{\a\a})^n {\cal{T}}_{0\supsuboverbrace[2n]{0\cdots0}\a_{s-2n-1}}
\ee
and after contracting with the deformation parameter 
\be\label{xiP}
s\, \xi_{\a_{s-1}}\Pi^{r \a_{s-1}} \sim -\,C\sum_{n=0}^{[s/2]}D_{1}(n)I_{1}(n)\binom{s-1}{2n}(n-1)\,\chi^{\supsuboverbrac[2n]{0\cdots0}\a_{s-2n-1}}{\cal{T}}_{0\supsuboverbrace[2n]{0\cdots0}\a_{s-2n-1}} \, ,
\ee
where $C$ is the same factor as in \eqref{B1}, while $I_{1}(n) = \binom{s-1}{2n}(2n-1)!!$ is a different degeneracy factor that takes into account the number of non-equivalent terms in $\Pi^{r\a_{s-1}}$ that give the same contribution after contraction with $\xi_{\a_{s-1}}$ --~owing to the complete symmetric character of the latter. Then by adding up (\ref{B1}) and (\ref{xiP}) one gets
\be
\lim\limits_{r\rightarrow\infty}Q_{1} = C \!\int\! d^{d-2}x\sum_{n=0}^{[(s-1)/2]} \!\binom{s-1}{2n}\! \left[ n -(n-1)\frac{2^{n}n!(2n-1)!!}{(2n)!}\right]\chi^{\supsuboverbrac[2n]{0\cdots0}\a_{s-2n-1}}{\cal{T}}_{0\supsuboverbrace[2n]{0\cdots0}\,\a_{s-2n-1}} \\
\ee
thus recovering \eqref{Qgen-chi1} thanks to the identity \mbox{$(2n)! = 2^n n! (2n-1)!!$} (and taking into account the trace constraints defined in \eqref{cons-curr-s} and \eqref{conf-kill}).

A similar analysis yields 
\be \label{Ptr}
\tilde{\Pi}^{r\alpha_{s-4}} \sim -\,\frac{C}{2}\,r^{-2(s-4)-1}\sum_{n=2}^{[\frac{s+1}{2}]}D_{2}(n)\binom{s-1}{2n-1}\frac{n-1}{s-2n+1}(g^{\alpha\alpha})^{n-2}({\cal{T}}^{[n]})^{\alpha_{s-2n}}
\ee
from where one obtains
\be \label{lambda-Pi}
\lambda_{i_{s-4}}\tilde{\Pi}^{ri_{s-4}}\sim C\sum_{n=2}^{[(s+1)/2]}\binom{s-1}{2n-1}D_{2}(n)I_{2}(n)\frac{(n-1)}{s-2n+1}\,\chi^{\supsuboverbrac[2n-1]{0\cdots0}\a_{s-2n}}{\cal{T}}_{\supsuboverbrace[2n]{0\cdots0}\,\a_{s-2n}} \, ,
\ee
In the expressions above we have introduced the combinatorial factors
\be
D_{2}(n) =\frac{2^{n-2}(n-2)!(s-2n+1)!}{(s-3)!} \, , \qquad
I_{2}(n) = \binom{s-4}{2n-4}(2n-5)!! \, ,
\ee
and \eqref{lambda-Pi} allows one to finally derive \eqref{Qgen-chi2}, again taking into account the trace constraints in \eqref{cons-curr-s} and \eqref{conf-kill}.

\end{appendix}

%%%%%%%%%%%%%%%%%%%%%%%%%%%%%%%%%%%%%%%%%%%%%%%%%%%%%%%%%%%%%%%%%%%%%
%%%%%%%%%%%%%%%%%%%%%%%%%%%%% REFERENCES %%%%%%%%%%%%%%%%%%%%%%%%%%%%
%%%%%%%%%%%%%%%%%%%%%%%%%%%%%%%%%%%%%%%%%%%%%%%%%%%%%%%%%%%%%%%%%%%%%

%\newpage

\end{document}